\documentclass[aps,prd,superscriptaddress,showpacs,nofootinbib,eqsecnum,amsfonts,amsmath]{revtex4-1}
\usepackage{epsfig}
\usepackage{longtable}
\usepackage{xcolor}
\usepackage{hyperref}

\usepackage{amssymb}
\usepackage{amsmath}
\usepackage[makeroom]{cancel}
\usepackage{slashed}

\def\nn{\nonumber}
\def\beq{\begin{equation}}
\def\eeq{\end{equation}}
\def\be{\begin{equation}}
\def\ee{\end{equation}}
\def\bea{\begin{eqnarray}}
\def\eea{\end{eqnarray}}

\def\la{\langle}
\def\ra{\rangle}
\def\ms{\overline{MS}}

\def\t{\tilde }
\def\Lam{\Lambda}

\def\ga{\gamma}
\def\ds{\displaystyle}

\begin{document}

\title{Chiral Condensate and Spectral Density at full five-loop and partial six-loop orders of 
Renormalization Group Optimized Perturbation}

\author{Jean-Lo\"{\i}c Kneur}
\author{Andr\'e Neveu}
\affiliation{Laboratoire Charles Coulomb (L2C), UMR 5221 CNRS-Universit\'{e} de Montpellier, 34095 Montpellier, France}

\begin{abstract}
We reconsider our former determination of the chiral quark condensate $\la \bar q q \ra$ 
from the related QCD spectral density of the Euclidean Dirac operator, using our Renormalization Group Optimized Perturbation
(RGOPT) approach. Thanks to the recently available {\em complete} five-loop QCD RG coefficients, and some other
related four-loop results, we can extend
our calculations exactly to $N^4LO$ (five-loops) RGOPT, and partially to $N^5LO$ (six-loops), the latter 
within a well-defined approximation accounting for all six-loop contents exactly predictable 
from five-loops RG properties.
The RGOPT results overall show a very good stability and convergence, giving 
primarily the RG invariant (RGI) condensate, 
$\la\bar q q\ra^{1/3}_{RGI}(n_f=0) = -(0.840_{-0.016}^{+0.020}) \bar\Lam_0 $,
$\la\bar q q\ra^{1/3}_{RGI}(n_f=2) = -(0.781_{-0.009}^{+0.019}) \bar\Lam_2 $,
$\la\bar q q\ra^{1/3}_{RGI}(n_f=3) = -(0.751_{-.010}^{+0.019}) \bar\Lam_3 $, 
where $\bar\Lam_{n_f}$ is the basic QCD scale in the $\ms$-scheme for $n_f$ quark flavors, 
and the range spanned is our rather conservative estimated 
theoretical error. This leads {\it e.g.} to  
$ \la\bar q q\ra^{1/3}_{n_f=3}(2\, {\rm GeV}) =  -(273^{+7}_{-4}\pm 13) $ MeV, 
using the latest $\bar\Lam_3$ values giving the second uncertainties.
We compare our results with some other recent determinations. As a by-product of our analysis 
we also provide complete five-loop and partial six-loop expressions of the perturbative QCD spectral density, 
that may be useful for other purposes. 
\end{abstract}
%
\maketitle
\section{Introduction}
The chiral quark condensate $\la \bar q q \ra$ is a 
main order parameter of spontaneous chiral symmetry breaking, $SU(n_f)_L\times SU(n_f)_R \to SU(n_f)_V$ 
for $n_f$ massless quarks. It is an intrinsically nonperturbative quantity, indeed 
vanishing at any finite order of ordinary perturbative QCD in the chiral limit. 
For nonvanishing quark masses, the famous Gell-Mann-Oakes-Renner (GMOR) relation~\cite{GMOR}, {\it e.g.} for the two lightest flavors:
\be
F^2_\pi\,m^2_\pi = -(m_u +m_d) \la \bar u u \ra +{\cal O}(m^2_q),
\label{GMOR}
\ee
relates the condensate with the pion mass $m_\pi$ and decay constant $F_\pi$ together with the (current) quark masses.
At present the light quark masses $m_{u,d,s}$ determined from lattice simulations (see \cite{LattFLAG19} for 
a recent review) 
give an indirect determination of the condensate from using (\ref{GMOR}). 
Phenomenological values of the condensate can also be extracted~\cite{qqSR,qqSRlast} 
indirectly from data using spectral QCD sum rule methods~\cite{SVZSSR}.
However, the GMOR relation (\ref{GMOR}) entails 
explicit chiral symmetry breaking from quark masses, and is valid up to  higher order
terms ${\cal O}(m^2_q)$. Thus more direct ``first principle'' determinations 
are always desirable to disentangle quark current mass effects for a better understanding 
of the dynamical chiral symmetry breaking mechanism at work in QCD. 
Analytical determinations have been derived in various models 
and approximations, starting early with the Nambu and Jona-Lasinio model~\cite{NJL,NJLrev}. 
There is also a long history of determinations based on Schwinger-Dyson equations
and related approaches\cite{DSErev,D-S,qqbar_link,DSEvar} typically. 
Lattice calculations have also determined the quark condensate by different approaches\cite{qqlatt_other}, 
in particular by computing the spectral density
of the Dirac operator\cite{qqlattSDearly,qqlattSD,SDlatt_recent}, directly related to the quark condensate via the
Banks-Casher relation\cite{BanksCasher,SDgen,SDchpt2}. Although some of the 
lattice determinations are very precise, those always rely on extra assumptions and modelization 
to extrapolate to the chiral limit\cite{CSBlatt_rev18}, using mainly chiral perturbation theory\cite{chpt}.
Moreover the convergence properties of chiral perturbation\cite{chptn3} for $n_f = 3$ are not as good 
as for $n_f=2$, and different recent lattice simulations 
still show rather important discrepancies\cite{LattFLAG19}. 
Also, within an extended chiral perturbation framework, it has been found significant 
suppression of the three-flavor case with respect to the two-flavor case~\cite{qqflav},
which may be attributed to the relatively large explicit chiral symmetry breaking from the strange quark mass. \\ 

Our renormalization group optimized perturbation (RGOPT)
approach~\cite{rgopt1,rgopt_Lam,rgopt_alphas} provides analytic sequences of  
(variational) nonperturbative approximations, having a non-trivial chiral limit.
As such it provides in particular an alternative independent determination of the chiral condensate\cite{rgoqq1}.
More generally the RGOPT method has also been explored so far in various models, in particular to improve 
the resummation properties of thermal perturbative expansions for 
thermodynamical quantities at finite temperatures\cite{rgopt_phi4},\cite{rgopt_nlsm}, and for QCD at  
finite densities~\cite{rgopt_qqmu}.
In the present work we iterate on our previous three- and four-loop RGOPT determination\cite{rgoqq1} 
of the condensate in the vacuum from the related spectral density, by going at the complete 
five-loop and partial six-loop level of our approximation.\\
 
In section II we shortly recall the well-known connection of the condensate with the spectral density of the Dirac 
operator through the Banks-Casher relation.
Also for completeness, in section III we shortly review our RGOPT variational construction of 
nonperturbative approximations, 
and its adaptation to the evaluation of the spectral density, as already detailed in ref.\cite{rgoqq1}. 
In section IV we derive the standard perturbative
quark condensate and related perturbative spectral density, exactly up to five-loop order and partially 
up to six-loop order in a well-defined approximation, 
thanks most notably to the recently available five-loop RG coefficients\cite{beta5l,gam5l}, in particular
the crucially relevant vacuum anomalous dimension~\cite{ga05l}. 
The perturbative spectral density for arbitrary number of quark flavors can also be useful 
for other purposes irrespectively of our
variational approach, most typically for perturbative matching of lattice simulation results. 
Section V give our detailed numerical analysis and the RGOPT condensate results order by order up to five
and (approximate) six loops, discussing also different approximation variants in order to estimate 
the theoretical uncertainties of our predictions.
In Section VI we compare with other recent determinations, mainly from lattice simulations. 
Finally section VII presents a summary and conclusions, and an Appendix completes various relevant 
expressions.
\section{Spectral density and the quark condensate}\label{spectral}
For a more detailed review of the connection of the density of eigenvalues $\rho(\lambda)$ of 
the Dirac operator with the chiral condensate $\la \bar q q  \ra$ through the Banks-Casher 
relation~\cite{BanksCasher}, we refer to our previous four-loop analysis~\cite{rgoqq1} and to former 
works and reviews (see {\it e.g.}~\cite{SDgen}).
The link between the spectral density and the condensate appearing in the operator product expansion (OPE) 
has been carefully discussed in \cite{qqbar_link}.
In short, in the infinite volume limit the spectrum of the Euclidean Dirac operator becomes dense, and 
using the formal definition of the quark condensate together with the properties of the eigenvalues of the 
Dirac operator leads to the relation
\be
\la \bar q q  \ra(m) = -2\, m \int_0^\infty {\rm d}\lambda \frac{\rho(\lambda)}{\lambda^2+m^2}\, .
\label{SD}
\ee
Eq.(\ref{SD}) essentially expresses that the two-point quark correlator 
has a spectral representation as a function of $m$.
The Banks-Casher 
relation is the chiral symmetric limit of Eq.\ref{SD}),
that gives the chiral condensate as
\be
\lim_{m\to 0} \la \bar q q  \ra = -\pi \rho(0)\, ,
\label{BC}
\ee
if the spectral density at the origin can be determined.
Note also that for non-zero fermion mass $m$, the spectral density is thus determined by the discontinuity of 
$\la \bar q q  \ra(m)$ across the imaginay axis:
\be
\rho(\lambda)=-\frac{1}{2\pi} \left[\la \bar q q  \ra ({\rm i}\lambda+\epsilon)-
\la \bar q q  \ra ({\rm i}\lambda-\epsilon)\right]|_{\epsilon\to 0}\, .
\label{disc}
\ee
For nonvanishing quark mass $m$, 
$\la \bar q q  \ra$ has a nontrivial perturbative series
expansion, $\sim m^3 \, f[\ln (m^2/\mu^2)]$, and its discontinuities are simply given
by those coming from the perturbative {\em logarithmic} mass dependence. Therefore the above relation (\ref{disc}) 
also allows to calculate the corresponding  perturbative spectral density. 
However, the $\lambda\to 0$ limit, relevant
for the true chiral condensate, trivially leads to a 
vanishing result, since perturbatively $\rho(\lambda)\sim \lambda^3$. 
But as we recall below a crucial feature of the variational RGOPT method
is to circumvent this,  giving a nontrivial result for $\lambda \to 0$.
\section{RG optimized perturbation (RGOPT)}\label{sec_rgopt}
\subsection{Optimized Perturbation (OPT) and RGOPT construction}
The RGOPT is basically a variational approach, made compatible with RG properties. 
The starting point is to deform the standard QCD Lagrangian by introducing a {\em variational} (quark) mass term, 
partly treated as an interaction term. One can most conveniently organize this
systematically at arbitrary perturbative orders, by introducing
a new expansion parameter $0<\delta<1$, interpolating between the (massive) free Lagrangian ${\cal L}_{free}$ and 
the  original (massless) Lagrangian ${\cal L}_{int}$ respectively.
This amounts first to the prescription:
\be m_q \to  m\:(1- \delta)^a,\;\; g \to  \delta \:g\;,
\label{subst1}
\ee
within some given (renormalized) perturbative expansion of a physical quantity $P(m,g)$
(here $g\equiv 4\pi\alpha_S$ for QCD). In Eq.(\ref{subst1}) 
we introduce for more generality an extra exponent $a$, 
that plays a crucial  role in our approach, as we recall below.  
Next the resulting expression is expanded in powers of $\delta$ 
at order $k$, the so-called $\delta$-expansion~\cite{delta}, and {\em afterwards} $\delta\to 1$ 
is taken to recover the original {\em massless} theory. This 
leaves a remnant $m$-dependence at any finite $k$-order: since at infinite $k$ order there is in principle 
no dependence on $m$, a finite-order approximation can be obtained through 
an optimization (OPT) prescription, {\it i.e.} a minimization of the dependence on $m$:
\be
\frac{\partial}{\partial\,m} P^{(k)}(m,g,\delta=1)\vert_{m\equiv \tilde m} \equiv 0\;,
\label{OPT}
\ee  
determining a nontrivial dressed mass $\t m(g)$. The prescription 
is consistent with renormalizability~\cite{gn2,qcd1,qcd2} 
and gauge invariance,  and (\ref{OPT}) realizes dimensional transmutation, 
in contrast with the original mass vanishing in the chiral limit.
In simpler one-dimensional models the procedure is a particular case of 
``order-dependent mapping''~\cite{odm}, and was shown to converge exponentially fast for the oscillator energy 
levels~\cite{deltaconv}. 

Now in most previous OPT applications, the simple (linear) value $a=1$ was used 
in Eq.~(\ref{subst1}) for the $\delta$-expansion mainly for simplicity.
In contrast we combine\cite{rgopt1,rgopt_Lam,rgopt_alphas} the OPT Eq.(\ref{OPT}) with renormalization group (RG)  
properties, by  requiring the ($\delta$-modified) expansion to satisfy, in addition to Eq.(\ref{OPT}), a 
perturbative RG equation:
\be
\mu\frac{{\rm d}}{{\rm d}\,\mu} \left(P^{(k)}(m,g,\delta=1)\right) =0, 
\label{RG}
\ee 
where the (homogeneous) RG operator acting on a physical quantity is defined 
as\footnote{Our normalization is $g\equiv 4\pi \alpha_S$, $\beta(g)\equiv 
{\rm d}g/{\rm d}\ln\mu$,  $\gamma_m(g) \equiv -{{\rm d} \ln m }/{{\rm d}\ln \mu}$,  see Appendix for relations
to ~\cite{beta5l,gam5l}.} 
\be
\mu\frac{{\rm d}}{{\rm d}\,\mu} =
\mu\frac{\partial}{\partial\mu}+\beta(g)\frac{\partial}{\partial g}-\gamma_m(g)\,m
 \frac{\partial}{\partial m}\;.
 \label{RGop}
\ee 
Note that once combined with Eq.~(\ref{OPT}), the RG equation takes a reduced massless form:
\be
\left[\mu\frac{\partial}{\partial\mu}+\beta(g)\frac{\partial}{\partial g}\right]P^{(k)}(m,g,\delta=1)=0\;.
\label{RGred}
\ee
Then a crucial observation is that after performing (\ref{subst1}),  perturbative RG invariance
is generally lost, so that Eq.~(\ref{RGred}) gives a nontrivial additional 
constraint~\footnote{A connection of the exponent $a$ with RG anomalous dimensions/critical exponents 
had also been established previously in the  $D=3$ $\Phi^4$ model for the Bose-Einstein condensate (BEC) 
critical temperature shift by two independent OPT approaches~\cite{beccrit,bec2}.}, but RG invariance can only be restored for 
a unique value of the exponent $a$, fully determined by
the (scheme-independent) first order RG coefficients~\cite{rgopt_Lam,rgopt_alphas}:
\be
a\equiv \ga_0/(2b_0)\;.
\label{acrit}
\ee
Therefore Eqs.~(\ref{RGred}), (\ref{acrit}) and (\ref{OPT}) together completely fix 
{\em optimized} $m\equiv \t m$ and $g\equiv \t g$ values. Moreover the prescription with (\ref{acrit})
drastically improves the convergence properties\cite{rgopt_alphas}.

Another known issue of standard OPT is that Eq.~(\ref{OPT}) alone 
generally gives more and more solutions as one proceeds to higher orders, with some being complex. Thus 
it may be difficult to select the right solutions, and unphysical (nonreal) ones are a burden. 
In contrast, the additional constraint (\ref{acrit}) guarantees that at arbitrary $\delta$ orders 
at least one of both the RG and OPT solutions 
$\t g(m)$ continuously matches  
the standard perturbative RG behavior for $g\to 0$ ({\it i.e.} Asymptotic Freedom (AF) for QCD):
\be
\t g (\mu \gg \t m) \sim (2b_0 \ln \frac{\mu}{\t m})^{-1} +{\cal O}( (\ln \frac{\mu}{\t m})^{-2}),
\label{rgasympt}
\ee
and these AF-matching solutions are often unique at a given order for both the RG and OPT equations.
However, (\ref{acrit}) does not guarantee in general that the compelling AF-matching solution remains 
real-valued for all physically relevant ranges. Actually the occurence of complex solutions is merely a consequence of  
solving exactly the (polynomial) Eqs.(\ref{OPT}, (\ref{RGred}), but since those equations
are derived from a perturbative expansion originally, they cannot be considered truly exact. 
Thus in practice one can often recover real solutions by considering a more approximate (perturbatively
consistent) RG equation or solution (see {\em e.g.} \cite{rgopt_alphas,rgopt_qqmu}). 
\subsection{RGOPT for the spectral density}
As shortly reviewed above in Sec.\ref{spectral}, using the spectral density with the Banks-Casher 
relation~(\ref{BC}) gives a direct access to the QCD condensate in the chiral limit.
Therefore the spectral density constitutes a particularly suitable Ansatz to apply our variational 
approach (see \cite{rgoqq1} for more discussions).
The RG equation relevant for $\rho(\lambda,a_s)$ was derived in \cite{rgoqq1} and is completely analogous to
the standard RG equation, but with the mass replaced by the spectral parameter, 
\be
\left[\mu\frac{\partial}{\partial\mu}+\beta(a_s)\frac{\partial}{\partial a_s}-\gamma_m(a_s)\,\lambda
 \frac{\partial}{\partial\lambda} -\gamma_m(a_s) \right]\rho(\lambda,a_s)=0.
\label{RGlam}
\ee 
One can next proceed to the modification of the resulting perturbative series $\rho(\lambda,a_s)$ 
as implied by the $\delta$-expansion, now, from Eq.~(\ref{RGlam}) clearly applied 
not on the original mass but on the
spectral value\footnote{We simplify 
notations with $\lambda\equiv|\lambda|$ since it is necessarily positive.} $\lambda$:
\be
\lambda\to \lambda (1-\delta)^a\, \;\;a_s\to \delta\,a_s\, .
\label{substlam}
\ee
Consequently the mass
optimization on $\la\bar q q \ra$ thus translates into an optimization
of the spectral density with respect to $\lambda$,
\be
\frac{\partial \rho^{(k)}(\lambda,a_s)}{\partial \lambda}= 0\, ,
\label{OPTlam}
\ee
at successive $\delta^k$ order (see \cite{rgoqq1} for more details).
 
Finally, as one last subtlety, note that the interpolation exponent $a$ in Eq.(\ref{acrit})
is universal in so far as the original expansion to be modified is itself (perturbatively) RG invariant. 
Now since $m \la \bar q q\ra$ is the RG invariant quantity, rather than $\la \bar q q\ra$, 
when performing the perturbative modification implied by (\ref{substlam}) on the spectral density, 
it is easily derived that the consistent value to be used is rather
\be
a=\frac{4}{3}(\frac{\gamma_0}{2b_0}),
\label{acritlam}
\ee
which also maintains the occurence of essentially unique 
AF-matching solutions with a behavior similar to (\ref{rgasympt})
(with $m\to \lambda$ understood).
\section{Perturbative quark condensate and spectral density}
\subsection{Perturbative quark condensate}
\begin{figure}[h!]
\epsfig{figure=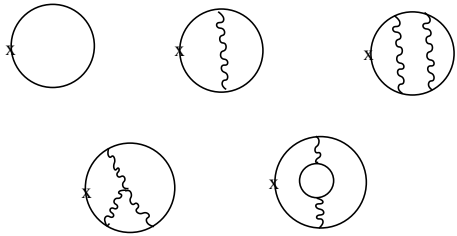,width=6cm}
\caption{Samples of standard perturbative QCD contributions to the chiral condensate 
up to 3-loop order. The cross denotes a mass insertion.}
\label{qqpert}
\end{figure}
The perturbative expansion of the QCD quark condensate for a nonzero quark mass can be calculated
systematically from the directly related vacuum energy graphs. 
A few representative Feynman graph contributions 
at successive orders are illustrated (up to three-loop order only) in Fig.~\ref{qqpert}. (There
are evidently some more three-loop contributions, not shown here). 
Note that the one-loop 
order is ${\cal O}(g^0)={\cal O}(1)$. 
The perturbative series for the renormalized
quantity $m \,\la \bar q q \ra$ up to six-loop order reads formally:
\bea
\displaystyle
& m\,\la \bar q q \ra(m,a_s) =
( \frac{3}{2\pi^2})\,m^4 \: 
\left \{ \frac{1}{2}-L_m +4 a_s (L_m^2 -\frac{5}{6} L_m +\frac{5}{12}) \right. \nn \\ 
& \left. +a_s^2 \sum_{i=0}^3 c_{3i} L_m^{3-i} + 
a_s^3 \sum_{i=0}^4 c_{4i} L_m^{4-i} + a_s^4 \sum_{i=0}^5 c_{5i} L_m^{5-i} 
+a_s^5 \sum_{i=0}^6 c_{6i} L_m^{6-i}  \right \}\, ,
\label{qqQCDpert}
\eea
where $L_m\equiv \ln (m/\mu)$, $m\equiv m(\mu)$ and $a_s\equiv \alpha_S(\mu)/\pi$ in the $\ms$ scheme
with renormalization scale $\mu$.
The two-loop contributions
were calculated in the $\ms$-scheme long ago, first in ~\cite{vac_anom2} (see also \cite{qcd2}).
At higher $k$-loop orders ($k\ge 3$) we have formally defined the  
coefficients as $c_{ki}$ for convenience, with their explicit expressions given below
and in the Appendix. Before detailing these expressions, 
we recall some rather well-known but important features related to RG properties. First, note  
that the calculation of the graphs in Fig. \ref{qqpert} still contains 
divergent terms, not cancelled by mass and coupling renormalization (as is clear already from the very first 
one-loop graph). Those divergences need an additive 
renormalization, in other words  $m \la \bar q q \ra$ 
has its own anomalous dimension, directly related to the (quark part of) 
vacuum-energy anomalous dimension. This also implies that the finite expression 
(\ref{qqQCDpert}) is not {\em separately} RG invariant: more precisely
the perturbative RG-invariance is expressed  in our normalization as
\be
\mu \frac{{\rm d}}{{\rm d}\,\mu} \left(  m\,\la \bar q q \ra(m,a_s) \right) + 4 m^4 \Gamma^0(a_s) \equiv 0 \; ,
\label{RGanom}
\ee
where the first term is the (homogeneous) RG operator given in Eq.(\ref{RGop}) and 
$\Gamma^0$ is the vacuum energy anomalous dimension\cite{vac_anom2,vac_anom3}, remarkably
recently evaluated fully analytically to five loops by the authors of Ref.\cite{ga05l} (see more details
in Eq.(\ref{Gam05l}) in Appendix).
Therefore note that the RG consistency expressed by requiring Eq.(\ref{RGanom}) to hold perturbatively order by order, 
allows to determine all the logarithmic $(L_m)^p$ coefficients
$c_{ki}$, with $k\ge i+2$ at perturbative orders $k$ from lower ($<k$) order coefficients 
and RG $\beta(a_s)$ and $\gamma_m(a_s)$ functions up to order $k-2$ and $k-1$ respectively. 
In addition the knowledge of $\Gamma^0(a_s)$ at $k$-loop order, together with lower-order
terms, fixes the remaining single logarithm coefficients $c_{k,k-1}$. The latter well-known RG properties 
constitute a crucial preliminary step of our RGOPT calculations, first requiring the
precise perturbative $m$-dependence, namely the relevant coefficients 
including massive quarks in (\ref{qqQCDpert}).\\
At three loops accordingly all the logarithmic coefficients $c_{3i}$, $i \le 2$ are easily determined~\cite{rgoqq1} as 
mentioned above from lower orders and RG properties. The 
remaining nonlogarithmic coefficient $c_{33}$, not related to RG properties, was calculated in \cite{qq3l}
from related three-loop quantities. 
In our normalization (and restricted to $N_c=3$ for QCD) these coefficients read: 
\bea
& c_{30} &= -\frac{2}{9}\,(81 -2\, n_f) , \nn\\
& c_{31} &= \frac{2}{9}\,(141 -5\, n_f ) ,\nn\\
& c_{32} &=  \frac{1}{16}\left( 52\, n_h +20 n_l -\frac{4406}{9} + \frac{32}{3}\, z_3 \right) , \nn\\ 
 & c_{33} &= \frac{1}{432}\left[ 6185 - 768\, a_4 - 32\ln^2 2 (\ln^2 2 -6 z_2) + 504 \,z_3  \right . \nn\\
&& \left.  + 528\, z_4  +  (672\, z_3 -750)\, n_h - 6 n_l (32 z_3 +45)  \right] ,
\label{c33}
\eea
for $n_l$ ``light'' (massless) and $n_h$ massive quarks, 
with $n_f=n_l+n_h$, $z_k =\zeta(k)$, and $a_4 = {\rm Li}_4(1/2)$.
In Eq.~(\ref{qqQCDpert}) $n_l$ and $n_h$ do not enter explicitly at one and two loops 
(fully described, up to unshown counterterms, by the first two graphs of Fig.\ref{qqpert}). 
At three loops $n_l$ and $n_h$ enter independently only within the $L_m$ 
and non-logarithmic coefficients $c_{32}$, $c_{33}$ respectively,
as can be deduced from inspection of the graphs of Fig.\ref{qqpert}.
To give a more numerical savor, in particular of the $n_l, n_h$ dependence and relative size compared
to the other (pure gauge) contributions,
one has to reasonable ($10^{-6}$) 
accuracy: 
\bea 
& c_{30} = -18 +\frac{4}{9}\, n_f \nn \\
& c_{31} = \frac{94}{3} -\frac{10}{9}\, n_f   \nn \\
& c_{32} = -29.7959 +3.25 n_h +1.25 n_l  \nn \\ 
& c_{33} = 16.4566 +0.133755\, n_h -1.15925 \, n_l \, .
\label{c3i}
\eea
Next at higher orders, using $\beta(a_s)$ and $\gamma_m(a_s)$ to four loops\cite{bgam4loop} and 
$\Gamma^0(a_s)$ to five loops\cite{ga05l}, we obtain after algebra the four-loop and five-loop exact analytical 
expressions of the logarithmic coefficients, given in Appendix (see Eqs.(\ref{c4k}), (\ref{c50})-(\ref{c54}). 
Numerically at four-loop order 
this reads~\footnote{We should point to a correction in $c_{43}$ here  as compared with
Eq.(5.13) of \cite{rgoqq1} (that was also differently normalized by an overall $4^3$ factor): 
this mistake, due to 
our previously incorrect interpretation of $n_h$ dependence from given $n_l=n_f -1$, $n_h=1$ results, 
changes $c_{43}$ by a few $0.1\%$, but  
affects our four-loop RGOPT condensate value by less than $10^{-3}$.}
\bea
& c_{40}& = 85.5 - 5.11111 n_f + 0.0740741 n_f^2 \nn \\
& c_{41}& = -224.333 + 16.7037 n_f - 0.246914 n_f^2 \nn \\
& c_{42}& = 342.151  -51.1008 n_h - 32.1008 n_l + 0.975309 n_h^2 + 0.308642 n_l^2 + 
+ 1.28395 n_h n_l  \nn \\
& c_{43}& = -375.082 + 42.6214 n_h + 43.5949 n_l - 0.0382074 n_h^2  - 
 0.790268 n_h n_l - 0.752061 n_l^2
\label{c4inum}
 \eea
and at five-loop order~\footnote{The authors of \cite{ga05l} provide  
the vacuum energy anomalous dimension at five loops for both diagonal contributions 
of $n_l$ massless and $n_h=1$ massive quark, and nondiagonal contributions ({\it i.e.} quarks of different masses).
It is straightforward to derive from their results the more specific 
case of $n_l=0$ and $n_h (=n_f)$ 
degenerate quarks of mass $m$, more relevant to our calculation.}
\bea
& c_{50}& = -418.95 + 42.1444 n_f - 1.38519 n_f^2 + 0.0148148 n_f^3 \nn \\
& c_{51}& = 1469.29 - 173.995 n_f + 6.01543 n_f^2 - 0.0617284 n_f^3 \nn \\
& c_{52}& =  -3079.72 + 436.666 n_f - 14.1506 n_f^2 + 0.102881 n_f^3 \nn \\
&& + n_h (155.167 - 
 11.7778 n_f  + 0.222222 n_f^2) \nn \\
& c_{53}& = 5102.45 - 852.446 n_h - 843.205 n_l + 61.7769 n_h n_l \nn \\
&& + 27.7449 n_h^2 + 34.032 n_l^2
- 0.344719 n_h^2 n_l \nn \\
&& - 0.701646 n_h n_l^2 
 + 0.00406919 n_h^3  -  0.352858 n_l^3 \nn \\
& c_{54}& =  ( n_f- 24.5)\, c_{44}
 -617.146  + 309.613 n_h + 144.324 n_l \nn \\ 
&& - 16.7381 n_h^2 - 4.8565 n_l^2
  - 21.5946 n_h n_l \nn \\
&& - 0.0719093 n_h^2 n_l  
- 0.0533908 n_h n_l^2 
- 0.0301426 n_h^3  - 0.0116241 n_l^3
\label{c5inum}
\eea
where we made clear the $c_{54}$ 
dependence upon the four-loop nonlogarithmic $c_{44}$ coefficient, not yet 
given explicitly at this stage as this deserves a more detailed discussion in the next subsection below.
Similarly we have derived all the six-loop coefficients that are determinable exactly from RG properties:
these are given in Appendix (see Eq.(\ref{c6k})).\\
Note that the nonlogarithmic five-loop
coefficient $c_{55}$ is presently not known, and this finite contribution (before renormalization) 
is presumably technically very challenging to evaluate. 
Fortunately it does not play any role in our (five-loop) determination below
since as above explained in Sec. \ref{spectral}, only the $\ln^p[m]$, $p\ge 1$ terms 
contribute to the spectral density, Eq.(\ref{disc}).
\subsection{Exact versus approximate determinations of $c_{44}$}\label{c44approx}
As just mentioned, we stress that more generally all the nonlogarithmic coefficients $c_{kk}$ in 
Eq.(\ref{qqQCDpert}) trivially do not contribute {\em directly} to the spectral density at any order $k$.
Yet these $c_{kk}$ are actually indirectly relevant, depending at which perturbative order one is performing 
calculations, since those coefficients enter
in the next order $c_{k+1\,k}$ {\em single} logarithm coefficient via RG properties, 
as explicited in Eq.(\ref{c5inum}) (see also Eq.(\ref{c54}) in Appendix).
The four-loop nonlogarithmic $c_{44}$ coefficient was not
known until very recently, nevertheless we could derive its approximate (but dominant) 
contribution, by exploiting other known four-loop results, as explained next. However while completing 
the present work, 
interestingly the complete $c_{44}$ has been very recently calculated\cite{maier0919}, which allows us to 
perform the five-loop RGOPT analysis with a fully known $c_{54}$ coefficient. \\
Let us first derive our approximation for $c_{44}$ (that we will also use in the numerics 
below, to assess the sensitivity
of our method upon such variations in the perturbative coefficients). 
For that purpose we exploit the relation of the condensate to another four-loop contribution as follows
\be
\frac{\partial}{\partial\,m} \left( \la \bar q q \ra(m)\right) = -\Pi_s(q^2=0) ,
\label{qq2PiS}
\ee
where $\Pi_s(q^2)\equiv i\,\int d^4 x e^{i q.x} \la 0|T J^s (x) J^s(0)|0\ra $ is the two-point scalar 
correlation function (the scalar current being defined as $J^s= \bar{q} q$).
This well-known relation (see {\it e.g.} \cite{chpt}) is valid to all orders
both at the bare and renormalized levels.  The various (vector, axial, (pseudo)scalar) correlators
have been investigated intensively in the literature~\cite{correl3lns,correl3ls}, and up to 
four loops~\cite{correl4l,PiSns4l}. 
In particular the  four-loop $\Pi_s(0)$ contribution was calculated in \cite{PiSns4l}, however not incorporating  
the so-called singlet contributions (as those were not directly relevant to the calculation of \cite{PiSns4l}).
(We recall that the singlet contributions, involving two disconnected quark lines in the two-point correlators, 
only appear starting at three-loop order, and for $\Pi_s(0)$ at three and four loops they are 
nonvanishing only for massive quark contributions $\propto n_h$). 
The {\em nonsinglet} four-loop nonlogarithmic contribution to $\Pi_s(0)$ is given in the $\ms$-scheme 
in Eq.(B.1) of ~\cite{PiSns4l}, a result that we recast here for completeness in our normalization conventions:
\be
\Pi_s^{4-loop, ns} (0) = (\frac{3}{2\pi^2})\, a_s^3 \, m^2 ( \frac{1}{2} \bar C_{-1}^{(3),s} +\ln (m/\mu)\, \mathrm{terms} ),
\label{c44sturm}
\ee
with
\bea
&&  \bar C_{-1}^{(3),s} = -325.6276432 +16.39537650 \,n_h +19.76434509 \,n_l \nn \\
&& -1.670198265\, n_h^2 - 0.9856898698 \, n_l n_h + 0.7103788267 \,n_l^2 .
\eea
Now at the level of the quark condensate, being a 
one-point function, there is no distinction between 'singlet' and 'nonsinglet' contributions, 
these being all included if the condensate is calculated from basics. 
But if deriving the condensate using Eq.~(\ref{qq2PiS}), we may explicitly 
separate the contributions that correspond to 'singlet' or 'nonsinglet' {\em within} $\Pi_s(0)$. 
Accordingly from a straightforward integration from Eq.(\ref{qq2PiS}) with 
input (\ref{c44sturm}), we obtain
the 'incomplete-singlet' (IS) approximation of $c_{44}$~\footnote{Eq.~(\ref{qq2PiS}) 
implies that $c_{43}$ also enter this relation.
Since $c_{43}$ is an exact contribution from the condensate,  
Eq.(\ref{c44IS}) involves both 'singlet' and 'nonsinglet' (from $\bar C_{-1}^{(3),s}$) contributions to
$\Pi_s(0)$.}: 
\bea
c^{IS}_{44} = && -\frac{1}{6} \bar C_{-1}^{(3),s} -\frac{1}{3} c_{43} \nn \\ 
= &&
179.2986813 -16.93968372 n_h + 0.2911021798 n_h^2  \nn \\
&&  - 17.82568131 n_l +  0.1322905474 n_l^2 + 0.4277044657 n_h n_l .
\label{c44IS}
\eea 
Note that the first (dominant) term in Eq.(\ref{c44IS}) 
is the pure gauge contribution, while terms $\propto n_l, n_h$ originate
from four-loop contributions with virtual massless and massive quarks.\\

Alternatively, the independent calculation very recently performed in 
\cite{maier0919} includes the complete contributions directly for the condensate: 
in the normalization of Eq.(\ref{qqQCDpert}) this full $c_{44}$ 
reads~\footnote{The original four-loop results of \cite{maier0919} 
combine exact analytical contributions with other (gauge) contributions known numerically but to 
very high accuracy of at least $10^{-74}$. 
Here we give for compactness the results numerically with $10^{-16}$ accuracy.}
\bea
& c_{44} & = 179.29868127533155 -15.013277376448457 n_h +0.7428868214454403 n_h^2 \nn \\
&& -17.825681312474572 n_l + 0.13229054734724904 n_l^2 + 
  1.0016895879838739 n_l n_h .
\label{c44ex}
\eea
As can be seen Eq.(\ref{c44IS}) 
is fully consistent with the complete result of Eq.~(\ref{c44ex}) 
(numerically within $10^{-10}$ relative accuracy) 
for its 'nonsinglet' part (including in particular
the dominant gauge contributions). 
Numerically the additional contributions within the full $c_{44}$  
are not at all negligible at four loops: for our relevant 
case with no massless quarks ($n_l=0$) and $n_h (=n_f)$ (degenerate) massive ones, 
Eq.(\ref{c44ex}) is $\sim 4\%$ ($\sim 7.5\%$) larger than 
(\ref{c44IS}), respectively for $n_f=2$ ($n_f=3$).
In the numerics below we evidently preferably use the full
Eq.(\ref{c44ex}), relevant for the five-loop spectral density via Eq.(\ref{c5inum}),
but in Sec. \ref{c44appxn} we also compare results obtained with the 'incomplete-singlet' approximation 
Eq.(\ref{c44IS}) in order to have a sensible estimate of the stability of five-loop RGOPT results with respect to this 
well-defined variation of the perturbative coefficients. We anticipate that it impacts the final
condensate value roughly by a $1 (2)\%$ change of the relative magnitude of $|\la \bar q q\ra|^{1/3}$ respectively
for $n_f=2$ ($n_f=3$).
\subsection{Explicitly RG invariant condensate}
One may use RG properties
to define a RG-invariant renormalized condensate expression, 
namely that obeys the homogeneous RG Eq.(\ref{RGop}), by compensating for the anomalous dimension in Eq.(\ref{RGanom}), 
as follows. 
The RG non-invariance of~(\ref{qqQCDpert}) can be perturbatively restored most simply upon considering 
perturbative extra {\em finite} subtraction contributions~\cite{qcd2,rgoqq1},
\be
\left[m \la \bar q q \ra \right]_{inv} \equiv m \la \bar q q \ra -S(m,a_s)  ,
\label{qq_sub}
\ee
where we define
\be
S(m,a_s) = \frac{3}{2\pi^2} \,\frac{m^4}{a_s} \sum_{k\ge 0} s_k  a_s^{k}\,
\label{sub}
\ee
with coefficients determined order by order by
\be
\mu \frac{\rm d}{{\rm d}\mu}{\rm S}(m,a_s) =\mu \frac{\rm d}{{\rm d}\mu}m (\la \bar q q \ra )\, = 
-4m^4 \Gamma^0(a_s) 
\label{rgsub}
\ee 
Once having determined as above all the correct logarithmic coefficients $c_{kj}$, $j <k$ at perturbative order $k$, 
one may apply the first equality in Eq.~(\ref{rgsub}), using the RG operator Eq.~(\ref{RGop}), to the 
finite expression~(\ref{qqQCDpert}), not separately RG invariant, to determine the subtraction function 
$S(m,a_s)$ uniquely. Of course, $S(m,a_s)$ actually only depends
on the vacuum energy anomalous dimension and other RG functions $\beta(a_s)$ and $\gamma_m(a_s)$, as the
second equality in (\ref{rgsub}) shows (which is nothing but a rewriting of Eq.~(\ref{RGanom}) above).
Note that Eq.~(\ref{sub}) necessarily 
starts with the $s_0/a_s$ term to be consistent with RG invariance properties.
In our normalization (\ref{sub}) the exact $s_i$  expressions up to five loops 
are given for completeness in Appendix A (see Eq.(\ref{s_i})).
Note, however, that Eq.(\ref{sub}) plays actually no role in our subsequent determination of the condensate
in the present work, since 
$S(m,a_s)$ does not involve any $\ln (m)$ terms, so trivially it does not contribute to the spectral density.
We have worked out this quantity for completeness since the
expression (\ref{qq_sub}) is nevertheless useful in other context (see e.g ref.\cite{rgopt_qqmu}). 
\subsection{Perturbative spectral density at five and six loops}
From the generic pertubative expansion for the condensate, 
Eq.~(\ref{qqQCDpert}),  calculating the (perturbative) spectral density 
formally involves calculating all logarithmic discontinuities according to Eq.~(\ref{disc}). 
This is simply given by taking in (\ref{qqQCDpert}) all non-logarithmic terms to zero, 
those having obviously no discontinuities,
while replacing all powers of logarithms, using $m\to {\rm i}|\lambda|$ etc.,  as 
\be
\ln^n \left(\frac{m}{\mu}\right) = \frac{1}{2^n} \ln^n \left(\frac{m^2}{\mu^2}\right)\to  \frac{1}{2^n} \frac{1}{2{\rm i} \pi}
\left[ \ln^n \left(\frac{|\lambda|^2}{\mu^2}\,e^{{\rm i}\pi}\right) -\ln^n \left(\frac{|\lambda|^2}{\mu^2}\,
e^{-{\rm i}\pi}\right) \right]\, ,
\label{discgen}
\ee
leading to the following substitution rules for the first few terms
\bea
&\ln \left(\frac{m}{\mu}\right) \to 1/2,\;\; 
\ln^2 \left(\frac{m}{\mu}\right) \to L_\lambda, \;\; 
\ln^3 \left(\frac{m}{\mu}\right) \to \frac{3}{2} L^2_\lambda -\frac{\pi^2}{8}, \nn \\
&\ln^4 \left(\frac{m}{\mu}\right) \to 2  L^3_\lambda -\frac{\pi^2}{2} L_\lambda ,\;\;
\ln^5 \left(\frac{m}{\mu}\right) \to \frac{5}{2} L^4_\lambda -\frac{5\pi^2}{4}  L^2_\lambda 
+\frac{\pi^4}{32} ,\;\; \nn \\
&\ln^6 \left(\frac{m}{\mu}\right) \to 3 L^5_\lambda -\frac{5\pi^2}{2}  L^3_\lambda 
+3\frac{\pi^4}{16}  L_\lambda,\;\;
\label{disc3}
\eea
where $L_\lambda\equiv \ln(\lambda/\mu)$
(note the $\pi^{2k}$ terms appearing starting at order $\ln^3 m$).\\
We obtain in this way the perturbative spectral density up to six-loop order formally:
\bea
& -\rho^{\ms}_{QCD}(\lambda,a_s)=  ( \frac{3}{2\pi^2}) \,\lambda^3 \:  
\left\{ -\frac{1}{2} + 4 a_s \,\left(L_\lambda -\frac{5}{12}\right) \right. \nn \\ 
& \left. +a_s^2 \sum_{i=1}^3 \rho_{3i} L_\lambda^{3-i} + 
a_s^3 \sum_{i=1}^4 \rho_{4i} L_\lambda^{4-i} + a_s^4 \sum_{i=1}^5 \rho_{5i} L_\lambda^{5-i} 
+ a_s^5  \sum_{i=1}^6 \rho_{6i} L_\lambda^{6-i} \right \}\, ,
\label{SD5l}
\eea
where the coefficients $\rho_{ki}$ for $k\ge 3$ are straightforwardly related
to the $c_{ki}$ of the original condensate using (\ref{disc3}) as follows:
\bea
&\rho_{31}& = \frac{3}{2} c_{30}, \nn \\ 
& \rho_{32}& =  c_{32}, \nn \\
&\rho_{33} & = \frac{1}{2} c_{32} -\frac{\pi^2}{8} c_{30}
\label{rho3i}
\eea
\bea
&\rho_{41}& = 2 c_{40}, \nn \\ 
& \rho_{42}& =  \frac{3}{2} c_{41}, \nn \\
&\rho_{43} & =   -\frac{\pi^2}{2} c_{40} +c_{42} \nn \\
&\rho_{44} & = -\frac{\pi^2}{8} c_{41} +\frac{1}{2} c_{43} 
\label{rho4i}
\eea
\bea
&\rho_{51}& = \frac{5}{2} c_{50}, \nn \\ 
& \rho_{52}& =  2 c_{51}, \nn \\
&\rho_{53} & =  -\frac{5}{4}\pi^2 c_{50} +\frac{3}{2} c_{52} \nn \\
&\rho_{54} & = -\frac{\pi^2}{2} c_{51} + c_{53} \nn \\
&\rho_{55} & = \frac{\pi^4}{32} c_{50} -\frac{\pi^2}{8} c_{52} +\frac{1}{2} c_{54}
\label{rho5i}
\eea
and so on at higher (six-loop) order (see Eq. (\ref{rho6k}) in Appendix).

At this stage, before proceeding with RGOPT, 
we remark that the above (ordinary) perturbative spectral density 
$\rho(\lambda)$ expression for arbitrary $n_l$, $n_h$ in Eq.~(\ref{SD5l}) 
can be useful for different purposes, independently of the 
RGOPT approach. For instance it should allow to proceed
at higher order the recently developed approach of ref.\cite{rho_latt_alphas}, 
to fit recent lattice precise calculations of the spectral density, 
in order to extract $\alpha_S$.
\section{Numerical RGOPT results for the condensate up to six loops}
We are now fully equipped to proceed with the main purpose, that we recap is  
to find solutions of the RGOPT equations, Eq.(\ref{OPTlam}), Eq.(\ref{RGlam}), applied to the spectral density
Eq.(\ref{SD5l}) at successive orders, after the modifications implied by Eq.(\ref{substlam}), 
(\ref{acritlam}). 
The RGOPT results up to four loops were obtained in \cite{rgoqq1} to which we refer for more details.
Here we will first summarize the main steps and important features for selfcontainedness, before
presenting in more details our new results at five and six loops. We also discuss in details the
numerical impact of some controllable approximations,  that will be specified, and how we
accordingly estimate theoretical uncertainties of our predictions.
\subsection{Summary of previous results up to four loops}
At one-loop order ${\cal O}(1)$ for the spectral density, (\ref{substlam}) only
affects the first constant term $-1/2$ in Eq.(\ref{SD5l}): since 
there is no logarithmic $L_\lambda$ contribution, 
one obtains for Eq~(\ref{OPTlam}) the trivial optimized solution, $\lambda = 0$. 
Thus nontrivial solutions occur starting at next-to-leading (NLO) two-loop order of the modified perturbation.
Accordingly at NLO order the modifed series reads
\be
-\rho^{\delta^1}_{QCD}=  \frac{3}{2\pi^2} \lambda^3 
\left(\frac{19}{58} +\frac{g}{\pi^2} ( L_\lambda -\frac{5}{12} )\right)\, ,
\label{rhod2l}
\ee
and the OPT~(\ref{OPTlam}) and RG~(\ref{RGlam}) equations
have a unique solution, 
given in the first lines of Tables~\ref{tabn2}, \ref{tabn3} for $n_f=2, 3$ respectively, using also~(\ref{BC}).
These results used the RG Eq.~(\ref{RGlam}) at one-loop order, that give simple analytic solutions. 
But since our optimized expression actually relies on exact two-loop
calculations, it appears more sensible to use the RG Eq.~(\ref{RGlam}) 
at the same (two-loop) order to incorporate {\it a priori}
more consistently higher-order effects. Doing this gives the results in the second lines of 
Tables~\ref{tabn2},~\ref{tabn3} for $n_f=2, 3$. 
Those results, to be considered more accurate, show a substantial decrease of 
the optimal coupling $\alpha_S$ to a more perturbative value
with respect to the results using the one-loop RG equation.

At higher orders the precise numbers obtained for the condensate also depend on the specific 
definition of the $\bar\Lam$ reference scale, which is generally perturbative 
and a matter of convention to some extent. The numbers in the first lines of Table \ref{tabn2} were obtained 
using the simpler one-loop form, $\bar\Lambda=\mu e^{-1/(2b_0 g)}$, consistently with the one-loop RG equation used. 
Next, when comparing below with other
determinations of the condensate, we use conventionally 
a four-loop definition of $\bar\Lam$ (see, eg., \cite{PDG}), in agreement with
most other past determination conventions. Except, at five-loop order, we obviously adopt 
the more consistent five-loop perturbative definition of   $\bar\Lam$.
The four-loop QCD scale $\bar\Lam$ expression reads in our normalizations (where $g\equiv 4\pi \alpha_S(\mu)$):
\be
 \ds \bar \Lam_{n_f}^{4-loop}(g) \equiv 
\mu  \,e^{-\frac{1}{2b_0\,g}}\,(b_0\:g)^{-\frac{b_1}{2b^2_0}}  \exp \left[ -\frac{g}{2b_0} \cdot  
 \left((\frac{b_2}{b_0}-\frac{b^2_1}{b^2_0})   
 +(\frac{b^3_1}{2b^3_0} -\frac{b_1 b_2}{b_0^2} +\frac{b_3}{2b_0}) g\right) \right]\;,
\label{Lam4pert}
\ee
with a straightforward generalization upon including the five-loop coefficient $b_4$. 
In Tables \ref{tabn2} and \ref{tabn3} we actually give for convenience the value of the  
scale-invariant condensate 
$\la\bar q q\ra_{RGI}$, which can be more appropriately compared between different perturbative orders. 
It is defined in our normalization as
\bea
&&\la\bar q q\ra_{RGI} \equiv \la\bar q q\ra(\mu)\;\exp{\left[\int\, {\rm d}g \,\frac{\gamma_m(g)}{\beta(g)}\, \right]} \nn \\
&&=  \la\bar q q\ra(\mu)\;(2b_0\,g)^{\frac{\gamma_0}{2b_0}} 
\left( 1+ (\frac{\gamma_1}{2b_0} -\frac{\gamma_0\,b_1}{2b^2_0})\,g 
+{\cal O}(g^2)\right)
\label{qqRGI}
\eea
where higher-order terms not shown here are easily derived, 
since only depending on the RG coefficients $b_i, \gamma_i$ known up to five loops. 
Remark, however, that our RGOPT optimization also fixes a scale, simply obtained from using
Eq.(\ref{Lam4pert}) (or its lower order equivalent) for $\bar \Lam(\t g)$, 
that are given indicatively in Table \ref{tabn2}, \ref{tabn3}. 
We stress that the optimal coupling 
$\t \alpha_S$ and corresponding optimal scale $\t \mu$, or the optimal spectral parameter $\t \lambda$, 
are to be considered intermediate values with no universal physical interpretation, since their precise
obtained values depend on the physical quantity being optimized. 
The physically meaningful result is obtained when inserting
$\t \alpha_S$ and $\t \lambda$ within the quantity being optimized, here $\rho(\lambda, \alpha_S)$. 
(This feature is quite general in optimization procedures: the values of the optimization parameters 
for a given physical quantity should not in general be used to evaluate another physical quantity).

At three-loop, $a_s^2$, order, the $n_f$ dependence appears explicitly within the perturbative coefficients of 
the spectral density, see Fig~\ref{qqpert} and the last 
$a_s^2$ coefficient in Eq.~(\ref{SD5l}). 
There occurs 
two real solutions for $\t L_\lambda, \t\alpha_S$, but the selection of the unique physical solution
is unambiguous since only one is clearly compatible
with AF behavior for $g\to 0$, $\ln (\t \lambda/\mu) \simeq -d_k/(2b_0 g) +{\cal O}(1)$ with 
$d_k={\cal O}(1)$, 
both for the RG and OPT equations.
In contrast the other real solution has for $g \to 0$ a coefficient 
of opposite sign to AF, and gives $\ln \t \lambda/\mu >0$, which is 
incompatible with perturbativity, since we expect $\mu \gg \t \lambda$ similarly to the perturbative
range $\mu \gg \t m \sim \bar\Lam$ for the original expansion with mass dependence. 
As stressed above in Sec. \ref{sec_rgopt} the occurence of an essentially unique solution with the correct 
AF-matching behavior at successive orders is a crucial feature of RGOPT, as will be illustrated further below.

At three and four loops the RGOPT results for $n_f=2, 3$ are specified in  
Tables~\ref{tabn2},~\ref{tabn3} respectively~\footnote{Since all the perturbative coefficients are known 
exactly at four loops, or to very high accuracy at five loops, our optimized results at a given order 
are in principle obtained to high accuracy. But in Tables~\ref{tabn2},~\ref{tabn3} (and similarly at higher orders
below) we give results to an accuracy largely sufficient for our purpose, given the extra 
uncertainties that will be discussed below.}. 
As indicated in each case  
we compare results obtained when using first the RG Eq.~(\ref{RGlam}) truncated at lower order,
and next taking the full RG equation at the same
three- or four-loop order respectively, incorporating more higher order dependence. 
At four-loops the raw optimization results actually give several real solutions for $\t \lambda, \t \alpha_S$ 
but there are no possible ambiguities since once more all solutions are eliminated from the AF-matching 
requirement, except a single one, with $\t \alpha_S >0$ and $\t L_\lambda <0$ as expected. 

One observes a further decrease of the optimal coupling $\t \alpha_S$ from three to four loops to more
perturbative values, as well as the corresponding decrease
of $\t L_\lambda$, meaning that $\t \mu$ is also larger.  The stabilization/convergence of the
results is clear for the scale-invariant condensate $\la\bar q q \ra_{RGI}$ given in 
Tables~\ref{tabn2},~\ref{tabn3}, which at four-loop order has almost no variation upon RG equation 
truncations~\footnote{In our numerical analysis below we use for convenience the exponentiated form of the
RG invariant factor as in (\ref{qqRGI}), but note that the relative difference with the fully perturbatively
expanded one is less than $10^{-3}$ for all considered optimized coupling values.}.
Note that the optimal
values $\t \alpha_S$ decreases substantially with increasing orders as compared to the lowest nontrivial order 
result above, thus indicating more perturbatively reliable results, moreover
$\t \alpha_S$ appears to somehow stabilize at three and four loops.
Notice also that compared with the more than 10\%
change in $\t \alpha_S$ upon going from two to three loops, 
the final physical condensate value only varies
by $0.25\%$, showing a strong stability. 
Also, while $\la\bar q q\ra^{1/3}/\bar\Lam$ changes by about $20\%$ compared to the
crude two-loop result in the first lines of Tables, it stabilizes rapidly at higher orders
showing {\it a posteriori} that the first nontrivial two-loop result seems  
already a quite realistic value. This stability at only NLO is a welcome feature 
for the usefulness of the RGOPT. A similar behavior was observed
when optimizing the pion decay constant in~\cite{rgopt_alphas}.
%
%
\begin{table}[h!]
\begin{center}
\caption[long]{$n_f=2$ RGOPT results at successive orders up to four loops for the spectral 
parameter $\t \lambda$, $\t \alpha_S$, and RG invariant condensate
 $\la\bar q q\ra^{1/3}_{RGI}$ calculated at the consistent perturbative order from~(\ref{qqRGI}). $\bar\Lam_2$
 is conventionally normalized in most cases by Eq.~(\ref{Lam4pert}), except in the very first line
 where the one-loop expression $\bar\Lam\equiv \mu\,e^{-1/(2b_0\,g)}$ is rather used. The 
corresponding scale values from Eq.~(\ref{Lam4pert}) are also given in the last column.
 } 
\begin{tabular}{|l||c|c|c|c|}
\hline
$\delta^k$, RG order  &  $\ln \frac{\t \lambda}{\mu} $ & $\t \alpha_S$ & 
 $\frac{-\la\bar q q\ra^{1/3}_{RGI}}{\bar\Lam_2}$ & $ \frac{\t \mu}{\bar\Lam_2}$ \\
\hline
$\delta$, RG 1-loop & $-\frac{2275}{10092}$ & $\frac{87\pi}{328}\simeq 0.83$  & $0.996$ &  $ 2.2$ \\
$\delta$, RG 2-loop &   $-0.45 $    & $0.480$ & $0.821$ & $ 2.8 $ \\
\hline
$\delta^2$,  RG 2-loop  & $ -0.686 $ & $ 0.483 $  & $0.792$ & $2.797 $\\
$\delta^2$,  RG 3-loop  & $ -0.703 $ & $ 0.430 $   & $0.783$  &   $3.104 $\\
\hline
$\delta^3$, RG 3-loop  & $-0.83895 $ & $0.40522$  & $0.77428$ &  $3.308$\\
$\delta^3$, RG 4-loop  & $-0.82164$ & $0.39071$ & $0.77247$  &  $ 3.448$\\
\hline
\end{tabular}
\label{tabn2}
\end{center}
\end{table}
\begin{table}[h!]
\begin{center}
\caption{Same captions as Table \ref{tabn2} for $n_f=3$} 
\begin{tabular}{|l||c|c|c|c|}
\hline
$\delta^k$ order  &  $\ln \frac{\t \lambda}{\mu} $ & $\t \alpha_S$ & 
$\frac{-\la\bar q q\ra^{1/3}_{RGI}}{\bar\Lam_3} $ & $\frac{\t \mu}{\bar\Lam_3}$ \\
\hline
$\delta$, RG 1-loop &   $-\frac{283}{972}$    & $\frac{27\pi}{104}\simeq 0.82$& $0.987$  & $2.35$ \\
$\delta$, RG 2-loop &   $-0.56 $    & $0.474$   & $0.789$ & $ 3.06$\\
\hline
$\delta^2$,  RG 2-loop  & $ -0.766 $ & $ 0.493 $   & $0.772$ & $2.942$ \\
$\delta^2$,  RG 3-loop  & $ -0.788 $ & $ 0.444 $   & $0.766$ &  $3.273$ \\
\hline
$\delta^3$, RG 3-loop  & $-0.97402 $ & $0.41367$  & $0.74377$  & $3.547$ \\
$\delta^3$, RG 4-loop  & $-0.96506$ & $0.39906$    & $0.74232$ & $ 3.709 $ \\
\hline
\end{tabular}
\label{tabn3}
\end{center}
\end{table}

\begin{figure}[h!]
  \centering
  \begin{minipage}[b]{0.4\textwidth}
    \includegraphics[width=6cm]{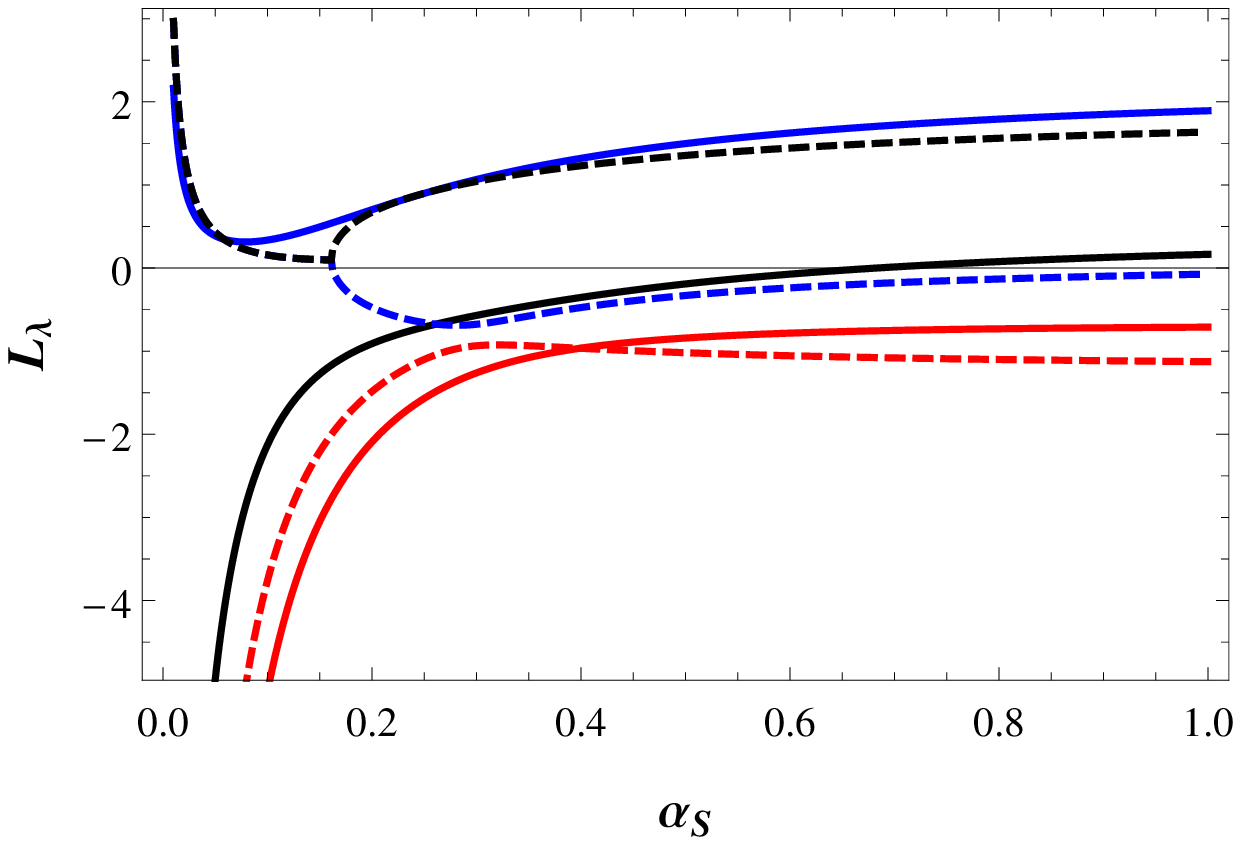}
  \end{minipage}
  \begin{minipage}[b]{0.4\textwidth}
    \includegraphics[width=6cm]{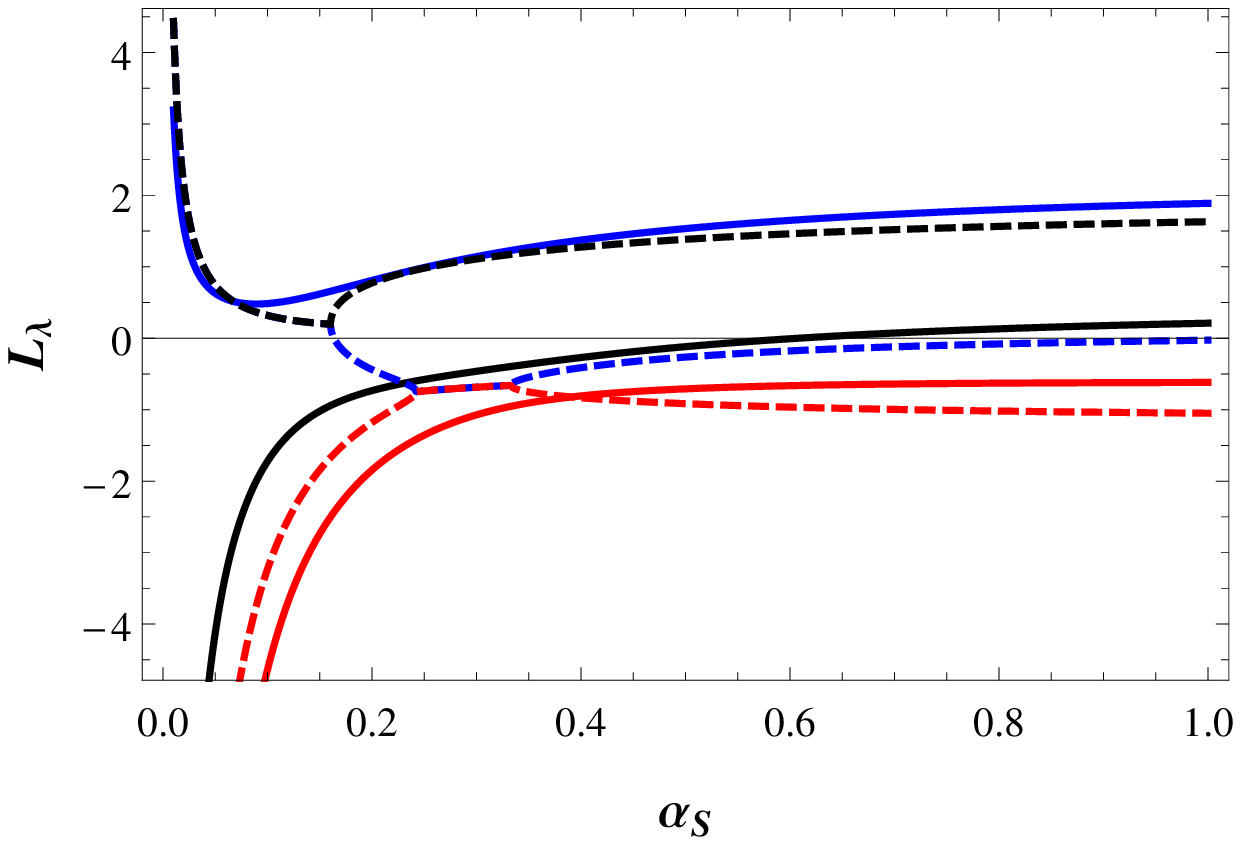}
  \end{minipage}
  \caption[long]{The different branch solutions $L_{\lambda}^{RG}(\alpha_S,k)$ (solid curves) and 
  $L_{\lambda}^{OPT}(\alpha_S,k)$ (dashed curves), $k=1,2,3$,  
  at $\delta^3$ (four-loop) order: left: $n_f=3$, right: $n_f=2$. Distinct branches appear to join within some 
$\alpha_S$ range where they actually become complex-conjugated, 
since only their real parts are plotted within this range.}
\label{branch4ln3n2}
\end{figure}
In Fig. \ref{branch4ln3n2}  we illustrate for $n_f=3$ (resp. left) and $n_f=2$ (resp. right) the 
different RG and OPT branches obtained at four loops, respectively as $L_{\lambda}^{RG}(\alpha_S,k)$, 
$L_{\lambda}^{OPT}(\alpha_S,k)$, where the number of solutions are at most $k=1,..3$ at four loops.
The clearest situation is the one for $n_f=3$,
where the two AF-matching RG and OPT branches (namely the two curves with the lowest $L_\lambda$ values
for any $\alpha_S>0$ in Fig. \ref{branch4ln3n2}) are real for any $\alpha_S >0$ and intersect at a unique
value, that determines unambiguously the solution, compare last line in Table \ref{tabn3}. 
Similar properties hold for all considered cases at lower orders. Next in Fig. \ref{branch4ln3n2} (right)
for $n_f=2$, two of the OPT branches unconveniently become complex (conjugate)-valued within 
the range $0.25 \lesssim \alpha_S \lesssim 0.33$ (so that their real parts shown here appear joined). 
Nevertheless one can still unambiguously 
select the correct AF-matching OPT branch, that is the one intersecting with the AF-matching RG branch, so
that the correct solution is again unique.
\subsection{Five-loop and six-loop results}\label{secVB}
Up to four loops, 
all the perturbative coefficients and RG quantities entering our evaluation have been known exactly
for some time.
Thanks to the recently calculated five-loop vacuum anomalous dimension\cite{ga05l}, Eq.(\ref{gam05l}), 
and the very recent complete calculation\cite{maier0919} of the four-loop nonlogarithmic 
coefficient $c_{44}(n_f)$, given in Eq.(\ref{c44ex}),
all the relevant perturbative coefficients needed at five loops are exactly available for 
the spectral density, Eq.(\ref{SD5l}).
Thus we can extend our evaluation for the physically relevant $n_f=2, 3$ values to five-loop order, 
correspondingly including up to five-loop contributions in the RG $\beta$ and $\gamma_m$ functions 
within the optimization, after performing consistently the $\delta$-expansion in Eq.~(\ref{substlam}) 
to order $\delta^4$. As mentioned above it is also useful to estimate the sensitivity of our results to the 
well-defined approximation Eq.(\ref{c44IS}) neglecting in $c_{44}$ the (subdominant) four-loop singlet contributions.
Furthermore, higher-order coefficients are (partly) determinable solely from perturbative RG invariance,
a feature that we can exploit to consider also (approximate) six-loop results (RGOPT order $\delta^5$). 
More precisely 
all the presently known five-loop RG coefficients, together with the complete four-loop coefficients, 
allow to determine {\em exactly} the six-loop coefficients 
$c_{6k}$, $0 \le k\le 4$ of  $\ln^{6-k} (m/\mu)$ of Eq.(\ref{qqQCDpert}). While the single 
logarithmic term ($k=5$) would need the presently unknown six-loop vacuum energy 
anomalous dimension as well 
as the five-loop nonlogarithmic coefficient $c_{55}$. 
The explicit expressions of the $c_{6k}$ are given in Eq.(\ref{c6k}) in Appendix.
Consequently from Eq.(\ref{disc3})
all the six-loop logarithmic terms, $\rho_{6k} L_\lambda^{6-k}$, $k=1,5$ of the spectral density 
in Eq.~(\ref{SD5l}) are exactly predicted (see Eq.(\ref{rho6k})), 
except for its last unknown nonlogarithmic coefficient. As we will examine below,
the six-loop results, although being approximate, are quite important to assess a more reliable 
determination of the condensate, due to the occurence of rather unwelcome instabilities for the strictly 
five-loop results.
\subsubsection{Five-loop and six-loop $n_f=2$ results}
\begin{figure}[h!]
\epsfig{figure=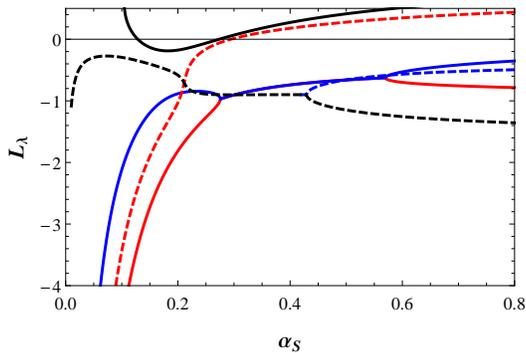,width=7cm}
\caption[long]{Some of the different RG (solid curves) and OPT (dashed curves) branches,
respectively $L_{\lambda}^{RG}(\alpha_S)$, $L_{\lambda}^{OPT}(\alpha_S)$
at $\delta^4$ (five-loop) order for $n_f=2$. Distinct branches appear to join within some 
$\alpha_S$ range where they actually become complex conjugate, 
since only their real parts are plotted within this range.}
\label{branch5ln2}
\end{figure}

We examine now in some details our procedure and results for $n_f=2$, with quite similar features  
given more briefly below for $n_f=3$, except when important differences need to be 
mentioned.
For $n_f=2$, at five loops we obtain one real solution that appears at first sight 
the closest to the lower (four-loop) results, namely:
$L_{\t \lambda} = -0.5699, {\t \alpha_S} = 0.5963$,  which gives 
$\la\bar q q\ra^{1/3}_{RGI} \simeq -0.863 \bar\Lam_2$, obtained using the RG equation at five-loops. 
(Very close results are obtained if using instead the RG equation at four loops). 
Without further inquiries one would conclude from this result  
that the five-loops RGOPT
produces an anomalously large shift of the condensate value, as compared with 
the seemingly well-stabilized three- and four-loop results $\sim -(0.78-0.77)$ in Table \ref{tabn2}. 
A directly related issue is the anomalously large optimized coupling that corresponds to this solution, 
$\alpha_S\sim 0.6$, in contrast with the regularly 
decreasing coupling obtained at increasing orders up to four loops, in Tables \ref{tabn2}, \ref{tabn3}. \\
However, upon applying our general criteria to select the correct solutions, a more careful examination 
shows that this solution cannot be correct, since it is not sitting on the perturbative 
AF-matching branch, in contrast to what occurs systematically at lower orders. 
This feature can be checked rather easily by perturbatively expanding at first order 
the four different branch solutions for RG and OPT Eqs. respectively, that both give quartic equations in $L_\lambda\equiv \ln (\lambda/\mu)$ at five-loops 
(thus respectively giving $L_\lambda^{RG}(\alpha_S,k)$, $L_\lambda^{OPT}(\alpha_S,k)$
with $k=1,..4$),
and examining which one(s) exhibit the perturbative AF-matching behavior, and whether 
the latter are matching the optimized $L_{\t \lambda}, \t \alpha_S$ values obtained at the 
intersecting solution(s). 
Equivalently it can be seen more pictorially in Fig. \ref{branch5ln2}, illustrating the different branches and 
(some of) their intersecting solutions (these branches are shown in a somewhat restricted but physically relevant range of 
$L_\lambda$ and $\alpha_S$): 
in contrast with the four-loop results in Fig. \ref{branch4ln3n2}, 
the RG branches now also become complex, similarly to the OPT branches, within a rather important $\alpha_S$ 
range: $0.27 \lesssim \alpha_S \lesssim 0.56$, and the only real intersection 
occuring at $\alpha_S\simeq 0.596, L_{\t \lambda}\simeq -0.57$ (visible near the top-right of Fig. \ref{branch5ln2}) 
sits on RG and OPT branches that are not linked to the correct AF behavior.\\

Therefore, at five-loop order the RG and 
OPT AF-matching branch do not have a real-valued
common intersection: a feature which somewhat complicates our investigation as compared with lower 
orders.
This large perturbation, sufficiently destabilizing the regular trend observed at lower orders to
suppress real AF-matching solutions, has two distinct, clearly identified origins. 
The first feature (but having a rather moderate impact on final results) is that 
the five-loop vacuum anomalous dimension coefficient~\cite{ga05l} is much larger 
relative to lower orders, moreover varying very much with $n_f$ (compare 
$\Gamma^0_4$ in Eq.(\ref{gam05l}) with $\Gamma^0_3$ in Eq.(\ref{gam04l}) in Appendix). 
(In contrast, as illustrated below,  
going from the four-loop to the five-loop $\beta$-function for the RG equation has a very modest impact, which
can be traced to the moderate numerical changes of adding five-loop RG coefficients to $\beta(\alpha_S)$,
$\gamma_m(\alpha_S)$ functions).
Note that $\Gamma^0_4$ enters the five-loop condensate $L_m$ coefficient $c_{54}(n_f)$, but it is 
not the sole contribution: the net effect from $\Gamma^0_4$ is typically that 
$|c_{54}(n_f)/c_{50}(n_f)| \sim 10$ roughly, which may be compared qualitatively 
with the similar four-loop quantities giving $|c_{43}/c_{40}|\sim 3$. 
But the second feature, that upon inspection happens to be the principal reason why the AF-matching solution 
is pushed into the complex domain, 
is that the discontinuities from Eq.(\ref{discgen}) entail also a relatively large term $\sim \pi^4$, appearing 
 for the first time at five-loop order in $\rho(\lambda,a_s)$, within the nonlogarithmic coefficient. 
More precisely, it is the last term of Eq.(\ref{rho4i}), modifying the relevant original perturbative
coefficient, $c_{54}(n_f)$ in Eq.~(\ref{qqQCDpert}) by about $60\%$, while all other
coefficients are more moderately affected by the discontinuity contributions. Simply ignoring this contribution
would be clearly inconsistent, 
and we will examine below how to better circumvent those problems.\\
Note on Fig. \ref{branch5ln2} that the AF-matching RG and OPT branches have not disappeared 
but just became complex (conjugate) valued within a
certain $\alpha_S$ range, rather unfortunately where the sought intersecting solution is expected.
Indeed one can determine precisely the complex-conjugated solution that sits on the AF-matching
branch: using the four-loop RG equation, we obtain: $L_{\t \lambda}\simeq -0.778 \pm 0.303 i, 
\t \alpha_S\simeq  0.358 \pm 0.0537 i$, that gives for the 
(RG invariant) condensate: $\simeq -(0.801 \pm 0.0195 i)\bar\Lam(2)$ (see Table
\ref{tabn2_5lrho} for more details).
Accordingly the correct AF-matching branch, although complex-valued in the relevant range, 
happens to give a corresponding condensate value with a small imaginary part, and with a 
real part in smoother continuity with 
the four-loop real solution. Also the corresponding (real part of the) optimal coupling $\t \alpha_S$ 
is more reasonably smaller than for the (wrong) naive real solution above. 
Very similar results are obtained if using rather the five-loop RG equation 
(see Table \ref{tabn2_5lrho}).\\

At this stage without further investigation one may just take the real part
as the physically relevant result, and interpret the imaginary parts as a rough 
estimate of the theoretical uncertainties of the results (although this is presumably 
not the best possible prescription to estimate the intrinsic uncertainties).
But given that the unwelcome occurence of nonreal solutions is only a consequence 
of solving {\em exactly} the RG and OPT polynomial equations in $L_\lambda$, 
and that it is seemingly not far from a real solution at five loops, one can more 
appropriately attempt to recover real solutions by a variant of the procedure. 
Accordingly a first possibility is simply to (perturbatively) approximate the sought optimized solutions at five loops.
Alternatively another possibility is to proceed to next (six-loop) order: at least this is possible 
in the approximation of neglecting the nonlogarithmic six-loop coefficient, being the only contribution 
not presently derivable from already known lower order results (as explained above at the beginning of 
SubSec. \ref{secVB}). Let us examine in turn those two possibilities.
\subsubsection{Perturbatively truncated five-loop RG solutions}
At five loops, instead of solving exactly the relevant RG and/or OPT optimization
Eqs.(\ref{RGlam}), (\ref{OPTlam}), one can consider 
more perturbative approximations,
as long as those remain consistent with the original perturbative order considered. Indeed
the RG Eq.(\ref{RGlam}) generates terms of formally higher order than five loops:
more precisely it is easy to see that at five loops Eq.(\ref{RGlam}) acting on 
the five-loop ($\alpha_S^4$) spectral density Eq.(\ref{SD5l}) involves up to $\alpha_S^{9}$ terms, due to the highest 
five-loop RG contributions $\propto b_4 \alpha_S^6$, 
$\gamma_4 \alpha_S^5$ respectively. 
But $b_4$, $\gamma_4$  appear 
first at order $\alpha_S^5$, $\alpha_S^4$ respectively.
Accordingly a presumably sensible procedure is to truncate\cite{rgopt_Lam} the RG equation, 
suppressing higher-order terms in $\alpha_S$ until possibly recovering a real 
common RG and OPT solution. 
At the same time if suppressing too many higher-order terms 
one loses the consistency with the RG content required at a given (here four- or five-loop) order.
A similar reasoning shows that the  next order six-loop 
RG coefficients, $b_5, \gamma_5$ (presently not known), would enter first respectively 
the $\alpha_S^6$, $\alpha_S^5$ coefficients of the RG equation.\\
Therefore it appears sensible to truncate any $\alpha_S^k$, $k\ge 6$ in the result of Eq.~(\ref{RGlam}), that would be
anyway affected by presently unknown higher orders. Further truncating the $\alpha_S^5$ term implies, however, losing
any dependence from the five-loop $b_4$ (while it still involves the five-loop $\gamma_4$ one).
Accordingly we found instructive to consider the effects of successive truncations, progressively
suppressing the highest $\alpha_S^{9}$ down to $\alpha_S^6$ (or even possibly $\alpha_S^5$) terms and comparing.
This is done below, with all results compiled in Table \ref{tabn2_5lrho} obtained by optimizing 
the spectral density $\rho(\lambda,\alpha_S)$ and keeping only the AF-matching branch solution (unique at 
a given order). \\
\begin{table}[h!]
\begin{center}
\caption[long]{ $n_f=2$ results at RGOPT $\delta^4$ (five-loops) and partial $\delta^5$
(six-loops) for the (RG invariant) condensate 
$\la\bar q q\ra^{1/3}_{RGI}/\bar\Lam_2$ and the corresponding
optimal values of the spectral parameter $\t \lambda$, coupling $\t \alpha_S$, and scale $\t \mu$, 
from optimizing $\rho(\lambda,\alpha_S)$ with Eqs. (\ref{OPTlam}), (\ref{RGlam}). 
We compare the results of (perturbatively consistent) successive RG equation truncations. 
$\bar\Lam_2$  is normalized by Eq.~(\ref{Lam4pert}) when the four-loop RG equation is used or by its five-loop 
extension when the five-loop RG equation is used. } 
\begin{tabular}{|l||c|c|c|c|}
\hline
RGOPT[$\rho(\lambda,\alpha_S)$] & & & & \\ 
$\delta^k$, RG order  &  $\ln \frac{\t \lambda}{\mu} $ & $\t \alpha_S$ & 
 $\frac{-\la\bar q q\ra^{1/3}_{RGI}}{\bar\Lam_2}$ 
 & $\frac{\mu}{\bar\Lam_2}$ \\
\hline
$\delta^4$, RG 4-loop (full)  & $ -0.77785 \pm 0.30316 i$ & $ 0.35785 \pm 0.053706 i$ & 
$0.80084 \pm 0.019516 i$ & $3.6418$ \\
$\delta^4$, RG 5-loop (full)  & $ -0.78815 \pm 0.31499  i$ & $ 0.35181 \pm 0.0501 i$ & 
$ 0.80104 \pm 0.019553 i$ & $3.7403 $ \\
$\delta^4$, RG 5-loop  ($\alpha_S^{\bcancel{k \ge 9}}$) & $ -0.80175 \pm 0.3081 i$ & $ 0.35384 \pm 0.043583 i$ &
$ 0.80101 \pm 0.019379 i$ & $3.7596 $  \\
\hline
$\delta^4$, RG 4-loop  ($\alpha_S^{\bcancel{k \ge 8}}$) & $ -0.81664 \pm 0.29248 i$ & $0.35958 \pm 0.036002 i$ &
$0.80033 \pm 0.018530 i$ & $3.7292$  \\
$\delta^4$, RG 5-loop  ($\alpha_S^{\bcancel{k \ge 8}}$) & $ -0.8151 \pm 0.28951 i$ & $ 0.36097 \pm 0.036464 i$ &
$ 0.8009 \pm 0.018641 i$ & $ 3.7094$  \\
\hline
$\delta^4$, RG 4-loop  ($\alpha_S^{\bcancel{k \ge 7}}$) & $ -0.83759 \pm 0.25140 i$ & $ 0.37539 \pm 0.024696 i$ &
$ 0.80048 \pm 0.015453 i$ & $ 3.5816 $  \\
$\delta^4$, RG 5-loop  ($\alpha_S^{\bcancel{k \ge 7}}$) & $ -0.83618 \pm 0.250 i$ & $ 0.37603 \pm 0.025177 i$ &
$ 0.80121 \pm 0.01559 i$ & $ 3.5728$  \\
\hline
$\delta^4$, RG 4-loop  ($\alpha_S^{\bcancel{k \ge 6}}$) & $ -0.93852 \pm 0.17354 i $ & $0.40216 \pm 0.01192 i $ &
$0.79914 \pm 0.0004024 i $ & $3.3298 $  \\
$\delta^4$, RG 5-loop  ($\alpha_S^{\bcancel{k \ge 6}}$) & $ -0.93840 \pm 0.17332 i$ & $ 0.40223 \pm 0.011871 i$ &
$ 0.79994 \pm 0.0004919 i$ & $3.3292  $  \\
\hline\hline
$\delta^5$, RG 5-loop (full)  & $-1.0846 $ & $ 0.32689$ & 
$ 0.77133 $ & $4.3737 $ \\
$\delta^5$, RG 5-loop  ($\alpha_S^{\bcancel{k \ge 6}}$) & $ -1.1422$ & $0.33778 $ &
$0.77260 $ & $4.1671 $  \\
\hline
\end{tabular}
\label{tabn2_5lrho}
\end{center}
\end{table}

From the $n_f=2$ results of Table \ref{tabn2_5lrho}, at five loops it appears not that easy to recover real solutions: 
upon truncating terms progressively starting from highest order ones, the results do not change much
at first, although there is a slow but clear 
decrease of the corresponding imaginary parts. Also, despite the not small $Im[ L_{\t \lambda} ]$ values,
the resulting condensate has much smaller imaginary parts, 
and real parts remain very stable, differing relatively only by ${\cal O}(10^{-3})$ for the different truncations. 
Similarly, for all cases there are tiny differences between the results using the four-loop or five-loop
RG equation. The $\t \alpha_S$ value are also
more reasonably perturbative, and close to the real four-loop results of Table \ref{tabn2}.
Truncating maximally the RG equation (namely by all terms $\alpha_S^{k\ge 6}$, but that still involves 
all the five-loop RG coefficients), 
the correct (AF-matching) solution has a tiny imaginary part, so that its real part 
may be considered reliable, giving $\la\bar q q\ra^{1/3}_{RGI} \simeq -(0.800 \pm 0.0005 i)\bar\Lam_2$. 
Note that if further truncating the RG equation, one not only loses the consistent RG content at five-loop order,
but the corresponding RG equation no longer gives any AF-matching branch.
\subsubsection{Approximate (partial) six-loop RGOPT}
As sketched above, the second 
alternative is to proceed at next (six-loop) order of Eq.(\ref{SD5l}),
with the $\ln^{6-k} (\lambda)$ coefficients given explicitly in Appendix (see Eqs.(\ref{c6k}), (\ref{rho6k})). 
The motivation, apart simply from the fact that most of the six-loop coefficients are 
readily exploitable from RG properties,
is that the discontinuities (\ref{discgen}) entail additional contributions $\propto \pi^4$
(see the last terms in Eq.(\ref{disc3})),
that tend to partially balance the instability triggered by $\pi^4$ discontinuity terms 
appearing first at five-loop order. At this point it is worth remarking that such features are 
generically expected from Eq.(\ref{discgen}): typically in \cite{rgoqq1} 
we have calculated the spectral density for the Gross-Neveu (GN) $O(N)$ model~\cite{GN} 
in the large-$N$ limit, to very high perturbative orders, that exhibits a clear pattern: 
the RGOPT solutions at increasing orders 
converge slowly towards the exact result (known for the GN model),
those solutions being destabilized each time novel $\pi^{2k}$ contributions
appear first, at increasing orders~\footnote{The GN spectral density 
exhibits at low orders even a more pronounced destabilization than for QCD,
because the (large-$N$) basic perturbative expansion of $\la \bar q q \ra_{GN}(m)$ in the $\ms$ scheme has
vanishing nonlogarithmic coefficients. Therefore, relative to zero, the large contributions generated by 
(\ref{discgen}) within $\rho(\lambda)$ are maximally destabilizing corrections.}. However, 
if keeping only {\em fixed} $\pi^{2k}$ terms (namely discarding $\pi^{2k+2}$, etc, terms appearing at 
higher orders), remarkably at sufficiently high fixed order all the $\pi^{2k}$ terms cancel, and 
the exact GN spectral density is obtained~\cite{rgoqq1}. 

For the QCD spectral density such exact cancellations are not expected, moreover 
obviously we are quite limited in trying to reach still higher orders. But inspired from these properties it is worth 
comparing two available successive orders (five and six loops), that actually rely on the same five-loop RG content, 
since as we recall, five-loop RG properties predict most of the six-loop coefficients of $\rho(\lambda)$ (all
except the nonlogarithmic one, $\rho_{66}$ in Eq.(\ref{SD5l}).
Accordingly one should keep in mind that it remains an approximation to the complete six-loop results, 
since $\rho_{66}$ involves the presently unknown six-loop vacuum anomalous dimension 
and the five-loop nonlogarithmic coefficient $c_{55}$. Therefore we simply set $\rho_{66}$ to zero in 
our numerics (neglecting also consistently the other nonlogarithmic contributions, generated at 
six loops from the 
discontinuities (\ref{disc3}). 
We will argue below that this approximation should moderately deviate from the complete six-loop results.\\

\begin{figure}[h!]
\epsfig{figure=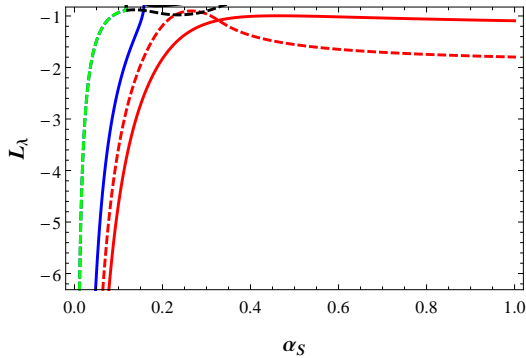,width=7cm}
\caption{Some of the relevant RG (solid curves) and OPT (dashed curves) branches,
respectively $L_\lambda^{RG}(\alpha_S)$, $L_\lambda^{OPT}(\alpha_S)$
at $\delta^5$ (6-loop) order for $n_f=2$. }
\label{branch6ln2}
\end{figure}
%
For $n_f=2$ the corresponding partial six-loop RGOPT results are given
in the last two lines of Table \ref{tabn2_5lrho}, also considering the (maximal) RG consistent truncation. 
As one can see a real solution is recovered at six loops, moreover the two AF-matching RG and OPT branches
remain real for all the physically relevant $\alpha_S$ range, and their intersection occur for a 
substantially smaller $\t \alpha_S$ value as compared to five loops.
This is illustrated also in Fig. \ref{branch6ln2}, zooming on the RG and OPT branches in the 
relevant range of $L_\lambda, \alpha_S$, which looks qualitatively more similar to the four-loop
$n_f=3$ case. It is striking that the resulting condensate value is much closer to the four-loop results,
that is not a numerical accident but is more essentially the effect of partially balancing at six loops
the instability from the large $\pi^4$ terms occuring first at five loops.  
\subsubsection{Summary of $n_f=2$ results}
As a tentative summary of the previous $n_f=2$ investigation:
\begin{itemize}
 \item 
At five loops, the impact of both large five-loop
vacuum energy anomalous dimensions and (more importantly) the first occurence of $\pi^4$ 
terms from (\ref{discgen}), are strong enough 
to destabilize the regular features observed at lower orders  up to four loops. 
Consequently one fails to obtain a strictly real AF-matching
solution.
Yet the five-loop results from successive truncations of unmandatory higher order
terms in the RG equation are very consistent,
reflecting a good stability.
Also the imaginary parts are small enough (especially for the maximal truncation of $\alpha_S^{k \ge 6}$, 
see Table \ref{tabn2_5lrho}) and can be included within the theoretical uncertainties.

\item Next, going to six loops restores a real unique AF-matching solution, that
results from a partial balance of the destabilizing $\pi^4$ terms. This solution  
has very regular properties and happens to be very close to the four-loop results.
\end{itemize}
These properties are more generically confirmed from comparison with the other relevant values $n_f=3$, 
or $n_f=0$, as illustrated next.
\subsection{Five- and six-loop $n_f=3$ results}
%
\begin{table}[h!]
\begin{center}
\caption[long]{$n_f=3$ results at five- and six-loops: same captions as in Table \ref{tabn2_5lrho}.} 
\begin{tabular}{|l||c|c|c|c|}
\hline
RGOPT[$\rho(\lambda,\alpha_S)$] & & & & \\ 
$\delta^k$, RG order  &  $\ln \frac{\t \lambda}{\mu} $ & $\t \alpha_S$ & 
 $\frac{-\la\bar q q\ra^{1/3}_{RGI}}{\bar\Lam_3}$ 
 & $\frac{\mu}{\bar\Lam_3}$ \\
\hline
$\delta^4$, RG 4-loop (full)  & $ -0.92148 \pm 0.29624  i$ & $  0.36925 \pm 0.051547 i$ & 
$0.76927 \pm 0.016561 i$ & $ 3.9124$ \\
$\delta^4$, RG 5-loop (full)  & $  -0.93013 \pm 0.3038  i$ & $ 0.36465 \pm 0.04859 i$ & 
$ 0.76838 \pm 0.016134 i$ & $3.9911  $ \\
\hline
$\delta^4$, RG 4-loop  ($\alpha_S^{\bcancel{k \ge 6}}$) & $ -1.064 \pm 0.13845 i $ & $ 0.42391 \pm 0.012618 i $ &
$0.77089 \pm 0.0024578 i $ & $3.4384 $  \\
$\delta^4$, RG 5-loop  ($\alpha_S^{\bcancel{k \ge 6}}$) & $ -1.0639 \pm 0.13833  i$ & $ 0.42394 \pm 0.012593 i$ &
$ 0.76975 \pm 0.0023702  i$ & $ 3.4297  $  \\
\hline\hline
$\delta^5$, RG 5-loop (full)  & $-1.2340 $ & $ 0.33863$ & 
$ 0.74042 $ & $4.6923 $ \\
$\delta^5$, RG 5-loop  ($\alpha_S^{\bcancel{k \ge 6}}$) & $-1.2618 $ & $ 0.3449$ &
$ 0.74076$ & $ 4.5578$  \\
\hline
\end{tabular}
\label{tabn3_5lrho}
\end{center}
\end{table}
For $n_f=3$ at five loops, very similarly to $n_f=2$ there is one real RGOPT solution appearing at first the closest
to the four-loop real results, using the RG equation at five loops:
$L_{\t \lambda} \simeq -0.693411$, $\t \alpha_S \simeq 0.598$. This gives for the
RG invariant condensate: $\simeq -0.813\bar\Lam_3$, thus a large shift from four-loop results of Table \ref{tabn3}.
But again upon examining the AF branches these do not match with this real solution, and the
correct AF-matching but complex-valued branch gives a more reasonable result, with small imaginary parts and 
real part closer to four-loop real results (see Table \ref{tabn3_5lrho}). \\
Similarly to $n_f=2$ we have performed a systematic analysis of all possible RG consistent truncations. 
We illustrate in Table \ref{tabn3_5lrho} the more relevant results, omitting intermediate case details. Overall the
behavior is much similar to $n_f=2$: at five loops one fails to recover strictly 
real AF-matching solutions, but the maximal truncation (discarding $\alpha_S^{k\ge 6}$, 
not loosing the five-loop RG content) gives very
small imaginary parts, and we will take the real part and add appropriate uncertainties in our final estimate. \\
Next, similarly to the $n_f=2$ results, at six-loop order one recovers a real AF-matching solution 
given in the last two lines in Table \ref{tabn3_5lrho},
with very regular properties and 
close to the four-loop results of Table \ref{tabn3}.
\subsection{Impact of approximated five-loop contributions}\label{c44appxn}
We now consider the approximation, defined in subsection \ref{c44approx} and 
relevant for $n_f=2, 3$, of using for the four-loop nonlogarithmic coefficient $c_{44}$
our expression in Eq.~(\ref{c44IS}) derived from the related nonsinglet four-loop 
scalar two-point correlator\cite{PiSns4l}. We recall that at the level of the 
optimized spectral density, this affects results only via the five-loop single logarithmic 
coefficient $c_{54}$. Since the previous results in subSec.\ref{secVB} 
including the very recently determined\cite{maier0919} exact $c_{44}$ 
coefficient Eq.~(\ref{c44ex}) are accordingly more complete, 
we will not include the variations resulting from this approximation within our uncertainty estimates. 
Nevertheless, given that the more exact five-loop results above are somewhat prevented by instabilities 
from producing real solutions, it is instructive to study their sensitivity upon such a 
well-defined approximation. 
The corresponding results are shown in Table \ref{tabn2n3appxn}
for $n_f=2$ and $n_f=3$ at five- and six-loops. 

Very similarly to the results obtained with the full $c_{44}$, at five loops the RGOPT 
gives nonreal AF-matching solutions, but with small imaginary parts.
Accordingly in Table \ref{tabn2n3appxn} the results are not too different from
the ones from using the exact $c_{44}$ in Tables \ref{tabn2_5lrho},  \ref{tabn3_5lrho}, except that 
the imaginary parts are somewhat smaller. For $n_f=3$, real AF-matching solutions are recovered
upon maximal truncations consistent with RG at five loops. 
At six loops real AF-matching solutions are also recovered, and these differ from the ones with exact $c_{44}$
in Tables \ref{tabn2_5lrho},  \ref{tabn3_5lrho} by about $\sim 1\%$ ($\sim 2\%$) lower in $\bar\Lambda$ units
for $n_f=2$ ($n_f=3$) respectively.
All these features are consistent with the fact that 
the loss of real solution at five loops is essentially due to the occurence of relatively 
large $\pi^4$ terms, while the $\sim 4\%$ ($\sim 7\%$) for $n_f=2$ ($n_f=3$) decrease 
in the approximated $c_{44}$ Eq.(\ref{c44IS}), as compared with the complete one Eq.(\ref{c44ex}),
has a more moderate impact.\\
We conclude that the approximation neglecting the four-loop singlet contributions within 
$c_{44}$ produces a change in the final condensate magnitude $|\la \bar q q \ra |^{1/3}$
that is about $\sim 1\% (2\%)$ smaller in magnitude respectively for $n_f=2$ ($n_f=3$), 
which again reflects a good overall stability.
\begin{table}[h!]
\begin{center}
\caption[long]{$n_f=2$ and $n_f=3$ results at five and six loops
using the approximate $c^{IS}_{44}$ coefficient from Eq.(\ref{c44IS}). 
Same captions as in Tables \ref{tabn3_5lrho}.} 
\begin{tabular}{|l||c|c|c|c|}
\hline
RGOPT[$\rho(\lambda,\alpha_S)$] & & & & \\ 
$\delta^k$, RG order  &  $\ln \frac{\t \lambda}{\mu} $ & $\t \alpha_S$ & 
 $\frac{-\la\bar q q\ra^{1/3}_{RGI}}{\bar\Lam_2}$ 
 & $\frac{\mu}{\bar\Lam_2}$ \\
\hline
$n_f=2$: & & & & \\
$\delta^4$, RG 4-loop (full)  & $ -0.71996 \pm 0.23156 i$ & $  0.37549 \pm 0.05979 i$ & 
$ 0.79341 \pm 0.0099350 i$ & $ 3.4177$ \\
$\delta^4$, RG 5-loop (full)  & $ -0.74047 \pm 0.24939   i$ & $0.36580 \pm 0.053884  i$ & 
$ 0.79353 \pm 0.0098374 i$ & $ 3.5523  $ \\
\hline
$\delta^4$, RG 4-loop  ($\alpha_S^{\bcancel{k \ge 6}}$) & $ -0.92679 \pm 0.10146 i $ & $ 0.39942 \pm 0.0074906 i $ &
$0.79090 \pm 0.0015884 i $ & $ 3.3590 $  \\
$\delta^4$, RG 5-loop  ($\alpha_S^{\bcancel{k \ge 6}}$) & $ -0.926667 \pm 0.10103  i$ & $ 0.39949 \pm 0.0074377 i$ &
$  0.79166 \pm 0.0016397 i$ & $ 3.3583   $  \\
\hline
$\delta^5$, RG 5-loop (full)  & $-1.15411 $ & $ 0.32270 $ & 
$0.75896 $ & $4.4605 $ \\
\hline
\hline
$n_f=3$: & $\ln \frac{\t \lambda}{\mu} $ & $\t \alpha_S$ & 
 $\frac{-\la\bar q q\ra^{1/3}_{RGI}}{\bar\Lam_3}$ 
 & $\frac{\mu}{\bar\Lam_3}$ \\ 
\hline 
$\delta^4$, RG 4-loop (full)  & $ -0.77002 \pm 0.19327  i$ & $ 0.40513 \pm 0.082108 i$ & 
$ 0.75375 \pm 0.0019729 i$ & $3.3255 $ \\
$\delta^4$, RG 5-loop (full)  & $ -0.79824 \pm 0.18251  i$ & $ 0.40213 \pm 0.067656 i$ & 
$ 0.75291 \pm 0.00097752 i$ & $ 3.4381  $ \\
\hline
$\delta^4$, RG 4-loop  ($\alpha_S^{\bcancel{k \ge 6}}$) & $  -1.1570 $ & $  0.42880 $ &
$ 0.76070 $ & $ 3.3996 $  \\
$\delta^4$, RG 5-loop  ($\alpha_S^{\bcancel{k \ge 6}}$) & $  -1.1572 $ & $ 0.42885 $ &
$  0.75957 $ & $ 3.3905  $  \\
\hline
$\delta^5$, RG 5-loop (full)  & $-1.3589 $ & $ 0.33097 $ & 
$ 0.71983$ & $ 4.8713$ \\
\hline
\end{tabular}
\label{tabn2n3appxn}
\end{center}
\end{table}

\subsection{The condensate in the quenched approximation}
One can also easily extend our calculations formally for $n_f=0$: 
actually a quark of mass $m$, setting the overall scale in Eq.(\ref{qqQCDpert}), 
'dressed' at higher orders by pure gauge interactions, 
is still understood in this case, and the perturbative removal of the quarks entering at three loops
in Fig.\ref{qqpert} and higher orders may be viewed as a perturbative analog 
of the 'quenched' approximation, which has its own theoretical interest, and can be compared 
with lattice simulations as we will examine.  
Specializing our calculations to the quenched approximation
from the known exact $n_f$ dependence in all relevant perturbative and RG coefficients,
one simply takes $n_f=0$ everywhere consistently.
Proceeding as previously described, from the first NLO (two-loop) nontrivial order up to five loops,
gives the results in Table \ref{tabn0}. \\
\begin{table}[h!]
\begin{center}
\caption{ $n_f=0$ (quenched approximation) results up to five loops: same captions as Table \ref{tabn2}. } 
\begin{tabular}{|l||c|c|c|c|}
\hline
$\delta^k$ order  &  $\ln \frac{\t \lambda}{\mu} $ & $\t \alpha_S$  & 
$\frac{-\la\bar q q\ra^{1/3}_{RGI}}{\bar\Lam_0}$ & $ \frac{\t \mu}{\bar\Lam_0}$ \\
\hline
$\delta$, RG 1-loop &   $-\frac{179}{1452}$ & $\frac{11\pi}{40}\simeq 0.86394$  & $1.0072$ &  $1.9370 $\\
$\delta$, RG 2-loop &   $-0.27548 $    & $0.49747$   & $0.9122$ & $ 2.4945 $ \\
\hline
$\delta^2$,  RG 2-loop  & $ -0.54201 $ & $ 0.47091 $   & $0.85679$ & $2.5795 $ \\
$\delta^2$,  RG 3-loop  & $ -0.54434 $ & $ 0.41691 $   & $0.83522$  &  $2.8224 $ \\
\hline
$\delta^3$, RG 3-loop  & $-0.63438 $ & $0.39314$  & $0.83741$  & $2.9713 $ \\
$\delta^3$, RG 4-loop  & $-0.60347$ & $0.38002$   & $ 0.83508$  & $ 3.0687 $ \\
\hline
$\delta^4$, RG 4-loop  & $-0.54077 $ & $0.39241$  & $0.85253 $ & $2.9764 $ \\
$\delta^4$, RG 5-loop  & $-0.45296$ & $0.42275$   & $0.86069 $ & $ 2.8045$ \\
\hline
\end{tabular}
\label{tabn0}
\end{center}
\end{table}
The solutions in Table \ref{tabn0} for the quenched case, up to five-loop order included, 
are all located on the real AF-matching branch (which is unique at a given order), although
when using five-loop RG in the very last line, the solution is located very close to the border of 
non-AF-matching branches. 
We observe that the condensate magnitude
$|\la\bar q q\ra|^{1/3}$ is driven to about $\sim 2.5\%$ higher values when going from four to five loops,  
as similarly observed above for $n_f=2, 3$ in Tables \ref{tabn2_5lrho}, \ref{tabn3_5lrho}. 
As above mentioned this is essentially traced to the instability from the first occurence of $\pi^4$ 
discontinuity term at five loops.\\
Although in this quenched case one obtains at five loops a real solution  upon using the complete
RG, it is also instructive to examine the trend obtained from
successive RG truncations, or alternatively when performing the calculation at six loops, giving
the results in Table \ref{tabn0_trunc}. 
\begin{table}[h!]
\begin{center}
\caption[long]{$n_f=0$ results at five and sixloops: same captions as in Tables \ref{tabn2_5lrho}.} 
\begin{tabular}{|l||c|c|c|c|}
\hline
RGOPT[$\rho(\lambda,\alpha_S)$] & & & & \\ 
$\delta^k$, RG order  &  $\ln \frac{\t \lambda}{\mu} $ & $\t \alpha_S$ & 
 $\frac{-\la\bar q q\ra^{1/3}_{RGI}}{\bar\Lam_0}$ 
 & $\frac{\mu}{\bar\Lam_0}$ \\
\hline
$\delta^4$, RG 4-loop (full) & $-0.54077 $ & $0.39241$  & $0.85253 $ & $2.9764 $ \\
$\delta^4$, RG 5-loop (full) & $-0.45296$ & $0.42275$   & $0.86069 $ & $ 2.8045$ \\
\hline
$\delta^4$, RG 4-loop  ($\alpha_S^{\bcancel{k \ge 6}}$) & $ -0.70122 + 0.11373 i $ & $ 0.36891 - 0.0031072 i $ &
$ 0.84724 + 0.00030466 i $ & $ 3.1610 $  \\
$\delta^4$, RG 5-loop  ($\alpha_S^{\bcancel{k \ge 6}}$) & $ -0.70089 + 0.11265 i$ & $  0.36906 - 0.0030291 i$ &
$ 0.85101 + 0.00019590  i$ & $ 3.170  $  \\
\hline\hline
$\delta^5$, RG 5-loop (full)  & $-0.90052 $ & $ 0.30873$ & 
$0.82242 $ & $ 3.912$ \\
$\delta^5$, RG 5-loop  ($\alpha_S^{\bcancel{k \ge 6}}$) & $-0.97183 $ & $0.31740 $ &
$0.82383 $ & $ 3.772$  \\
\hline
\end{tabular}
\label{tabn0_trunc}
\end{center}
\end{table}
As one can see here it is the effect of RG truncation that pushes the AF-matching solution to (slightly) 
nonreal values. 
Comparing these RG-truncated results from Table \ref{tabn0_trunc}, which have negligible imaginary parts, 
with the corresponding real solutions using the complete RG in Table \ref{tabn0},
gives a useful estimate of the impact of such RG-consistent truncations.
Concerning the six-loop results, very similarly to the $n_f=2$ and $n_f=3$ cases they are real 
and very regular, 
and again  much closer to the four-loop results in Table \ref{tabn0}.\\

Finally for completeness we have also considered the $n_f=1$ case: although it is not very relevant physically, 
it can be viewed at least as a further consistency crosscheck of our results.
We have explored variants similarly to other $n_f$ values above but  
simply summarize here the main results.
At four-loop order one obtains the unique real solution: 
\be 
\la\bar q q\ra^{1/3}_{RGI}(n_f=1,\mbox{4-loop})=-0.8039 \bar \Lam_1
\ee
which
appears very close to the 'average' of $n_f=0$ and $n_f=2$ four-loop results.  
At five-loops, using four- or five-loop RG, the real solution is no longer on the AF-matching branch,
similarly to the $n_f=2,3$ cases. The unique AF-matching solution obtained from 
truncating $\alpha_S^{k\ge 6}$ in the RG equation  gives:
\be
 \la\bar q q\ra^{1/3}_{RGI}(n_f=1,\mbox{truncated RG 5-loop})= -(0.8271 \pm 0.0007\, i)\bar \Lam_1 
\ee
Finally at six-loops a real solution is recovered, giving 
\be
 \la\bar q q\ra^{1/3}_{RGI}(n_f=1,\mbox{RG five-loop}) = -0.7984\bar \Lam_1 .
 \ee
Similarly to $n_f=0,2,3$ cases, once more the five-loop results produce a substantial 
$\sim 2.8\%$ increase of the condensate as compared
to four-loops results, 
while the six-loop results are very close to the latter.
\subsection{Evaluating theoretical uncertainties}
Comparing the different above results from $n_f=0$ to $n_f=3$, it is tempting to consider the manifestly more stable  
results obtained at four loops and six loops as a likely better approximation than the 
more sensibly shifted five-loop results. Note indeed that if discarding the latter, the combined 4-loop and 6-loop 
results would provide a seemingly very accurate determination.
But since the six-loop results are only partial, 
we more conservatively combine all those
results within our estimate of uncertainties. 
More precisely we take the average between the four-, five- and six-loop results as our central values 
and their differences as our theoretical uncertainties 
(taking at five loops the real parts of the results having the smallest imaginary parts, that are 
consequently more reliable). 
Then we estimate the uncertainties linearly
from the complete range spanned by maximal and minimal values.

For $n_f=0$, for which real solutions occur at all RGOPT successive orders considered, we obtain
\be
\la\bar q q\ra^{1/3}_{RGI}(n_f=0)\simeq  
-(0.840_{-0.016}^{+0.020})\bar \Lam_0,
\label{finn0}
\ee
where we give only a three digits accuracy given the uncertainties.

For $n_f=2$, proceeding similarly  we obtain
\be
\la\bar q q\ra^{1/3}_{RGI}(n_f=2)\simeq  
-(0.781_{-0.009}^{+0.019}) \bar \Lam_2 .
\label{finn2}
\ee

And finally for $n_f=3$:
\be
\la\bar q q\ra^{1/3}_{RGI}(n_f=3)\simeq  
-(0.751_{-.010}^{+0.019} ) \bar \Lam_3 .
\label{finn3}
\ee 
Eqs~(\ref{finn0}),(\ref{finn2}), (\ref{finn3}) constitute our primary results, as these do not depend
on any extra theoretical or experimental input besides the basic perturbative content used in the calculation
and the RGOPT method.   
Now, to  make contact with other independent determinations of the quark condensate, 
often conventionally given at the standard scale $\mu\simeq 2$ GeV for reference, 
one needs to perform a (perturbative) renormalization scale evolution. One should keep in mind that, 
in contrast with the above results,
such RG evolution unavoidably also entails $\alpha_s$ (and other related) uncertainties.
\section{ $\la\bar q q\ra (\mu=2\, \rm GeV)$ and comparison with other determinations}
To evolve perturbatively  
the condensate from our results above, the simplest
procedure is to take the values obtained for the scale-invariant condensate~(\ref{qqRGI}), within uncertainties,
Eqs.(\ref{finn0})-(\ref{finn3}) and extract from these
the condensate at another chosen (perturbative) scale $\mu'$, using again~(\ref{qqRGI}) at five-loop order,
now taking $g \equiv 4\pi\alpha_S(\mu')$, after evolving $\alpha_S(\mu)$ at five-loop order 
of Eq.(\ref{Lam4pert}) towards the conventional scale $\mu'=2$ GeV. The overall reliability of this (perturbative) 
evolution is to be assessed on the ground that the primary RGOPT results above 
at four-, five- and six-loops are obtained at reasonably perturbative optimized scale values 
($3.3  \Lambda_{\ms} \lesssim \t \mu \lesssim 4.8 \Lambda_{\ms}$ 
(compare Tables \ref{tabn2}-\ref{tabn3_5lrho}). It is more appropriate to separate the discussion below 
for different $n_f$ values, since those 
do not have all the same reliability status (also when comparing our results with other independent determinations
of the condensate) as we discuss next. We consider successively $n_f=3$, $n_f=2$, 
and $n_f=0$ (quenched approximation).
\subsection{ $n_f=3$ }
For $n_f=3$, one can use very reliable $\alpha_S$ determinations in the perturbative range. 
We also account properly for the charm
quark mass threshold effects\cite{matching4l}  on $\alpha_S(\mu\sim m_c)$.
From the most recent world average $\alpha_S(m_Z)$ value~\cite{PDG}:
\be
\alpha_S(m_Z)= 0.1179 \pm 0.0010 ,
\label{aswa19}
\ee
we obtain in a first stage, accounting for threshold effects at $\mu\sim m_b$ and 
$\mu\sim m_c$~\footnote{For the rather low values of the scales involved, it appears more appropriate to use
the exponentiated forms Eqs~(\ref{Lam4pert}), (\ref{qqRGI}), somewhat more stable than 
their purely perturbative expansions. We have also crosschecked
our five-loop RG evolution with the results using the well-known public code 
RunDec~\cite{QCDruncode}, recently upgraded to five-loop order.}:
\be
\bar\Lam(n_f=3) = (331 \pm 16)\, \mbox{MeV}
\label{Lam3}
\ee
and 
\be
\alpha_S(2\, \mbox{GeV}) = 0.3007 \pm 0.008 . \;\;
\label{aswa2GeV}
\ee
Then using Eq.~(\ref{qqRGI}) applied to Eq.(\ref{finn3}) leads to
\be
 \la\bar q q\ra^{1/3}_{n_f=3}(2 \, {\rm GeV}) =  -(0.826_{-.011}^{+.021}) \bar\Lam_3\;.
\label{qqn32GeV}
\ee
Thus combining Eq.(\ref{Lam3}) with (\ref{qqn32GeV}) leads to
\be
\la\bar q q\ra^{1/3}_{n_f=3}(2 \,{\rm GeV},\rm \bar\Lam^{wa}_3)\simeq  -(273^{+7}_{-4} \pm 13) \,{\rm MeV}\, 
\label{qqfin3wa}
\ee
where the first error is our rather conservative theoretical RGOPT uncertainty 
from Eq.(\ref{qqn32GeV}) and the second one 
is from $\bar\Lam(3)$ uncertainty. 
(Since these two uncertainties have very different origin we do not combine them).
It is worth remarking at this point that Eq. (\ref{qqn32GeV}) is 
only slightly shifted with respect to our previous (average of three- and four-loop RGOPT) result~\cite{rgoqq1}, 
while the central value and uncertainties in Eq.(\ref{qqfin3wa}) (compare Eq.(6.6) of \cite{rgoqq1}) 
are principally affected by the slight decrease of the most recent $\alpha_S$ world average
with substantial {\em increase} of uncertainties (see \cite{PDG} for detailed 
explanations on these features).\\
To compare with other independent determinations, first
the most precise $n_f=3$ lattice determination we are aware of, 
in the chiral limit, is $\la\bar q q\ra^{1/3}_{n_f=3}(2\rm GeV) =
-(245\pm 5 \pm 8)$ MeV~\cite{Lattqqn3}. Our results are thus marginally compatible with the latter, within 
uncertainties of both results. Note, however, that various recent lattice results vary in a wider
range for $n_f=3$, as compiled from \cite{LattFLAG19}:  
from $214\pm 6\pm 24$~\cite{Lattqqn3_low} to $ 290\pm 15$~\cite{Lattqqn3_high}. 
This is largely due to the still difficult required extrapolation of lattice results to
the chiral limit, which for the $SU(3)$ case is affected by large uncertainties. 
A recent very precise $n_f=3$ lattice calculation\cite{Lattqqn3_corr}, 
using time-moments of heavy-strange pseudoscalar 
correlator, has obtained $\la\bar s s\ra^{1/3}(\mbox{2 GeV}) \sim -(296\pm 11)$ MeV.
Since it is not in the chiral limit, it should not be {\em directly} compared with our result, given the 
large strange quark mass involved. Indeed
as our $n_f=3$ results are based 
on a relatively accurate RGOPT determination (\ref{finn3}), and (\ref{qqfin3wa}) obtained from a reliable 
$\bar\Lam(3)$ world average, they  
appear useful independent determinations since being in the strict chiral limit, thus relevant to possibly assess
the actual impact from explicit chiral symmetry breaking by the strange quark mass, 
by comparison with other determinations that include the latter, like \cite{Lattqqn3_corr}.
\subsection{ $n_f=2$ }
For $n_f=2$, one cannot directly link our results to the true phenomenological 
perturbative range values of $\alpha_S$ as above. 
Nevertheless, given that the (optimized) coupling values obtained in Table \ref{tabn2}, \ref{tabn2_5lrho} 
are reasonably perturbative, we can consider a perturbative (five-loop) RG evolution
(consistently performed in a simplified QCD picture where the strange and heavier quarks are all 
infinitely massive, {\it i.e.} `integrated out'). 
To give a final (numerical) determination of the condensate, we need a value for $\bar\Lam(n_f=2)$.
To our knowledge there are not so many nonperturbative results for $\bar\Lambda(2)$ (as compared with the 
numerous studies for $n_f=3$), 
and those results mostly originate from lattice calculations.
We therefore rely on a lattice determination\cite{Lam2_latt_ALPHA} 
(that best fulfill the reliability criteria of the review \cite{LattFLAG19}), 
obtained from the Schr\"odinger functional method:
\be
\bar\Lam(n_f=2) = (310\pm 20)\,\mbox{MeV} .
\label{Lam2}
\ee
One should keep in mind, however, that somewhat larger uncertainties are obtained if taking more conservatively
all presently available lattice results\cite{Lam2_latt_ALPHA,Lam2_latt_other}, as compiled in 
\cite{LattFLAG19}~\footnote{Our own determination
of $\bar\Lam(n_f=2)$ from the pion decay constant $F_\pi$ using three-loop RGOPT\cite{rgopt_alphas} 
is compatible with the range obtained from lattice simulations, but
also has larger uncertainties, as compared to our RGOPT $\bar\Lam(n_f=3)$ results.}.
After RG evolution up to 2 GeV we obtain accordingly from Eq.(\ref{finn2}):
\be
 \la\bar q q\ra^{1/3}_{n_f=2}(2 \, {\rm GeV}) =  -(0.863_{-.010}^{+.021}) \bar\Lam_2 . 
 \label{qqn22GeV}
\ee
(Notice that this strictly $n_f=2$
result, thus with the strange and heavier quarks integrated out, correspondingly has $\alpha_S(\mu)$ 
values not consistent with the phenomenological values of Eq.(\ref{aswa2GeV}): instead we find 
$\alpha_S^{n_f=2}(\mbox{2 GeV})\simeq 0.262\pm .008$).
Combining (\ref{qqn22GeV}) with Eq.~(\ref{Lam2}) leads to~\footnote{As compared 
with our 2015 four-loop result~\cite{rgoqq1}),
note that the central value in Eq.~(\ref{qqfin2}) is principally
affected by the somewhat lower central $\bar\Lam(n_f=2)$ value from (\ref{Lam2}) as compared with previously 
used value from \cite{LamlattVstatic14}.}
\be
\la\bar q q\ra^{1/3}_{n_f=2}(2 {\rm GeV},\rm \bar\Lam_2)\simeq  -(267^{+7}_{-4} \pm 18) \,{\rm MeV}\,, 
\label{qqfin2}
\ee
where again the first error range is our RGOPT uncertainty from Eq.(\ref{qqn22GeV}) while
the second one is from the $\bar\Lam(2)$ uncertainty. Since lattice uncertainties are
mostly statistical and systematic, while ours are theoretical, 
it is not obvious to combine these in a sensible manner and we keep more conservatively 
separate uncertainties.\\
To compare our result (\ref{qqfin2})
with other recent determinations, first the presumably most precise $n_f=2$ lattice determination 
to date is also from the spectral density~\cite{SDlatt_recent}:
$\la\bar q q\ra^{1/3}_{n_f=2}(\mu=2 {\rm GeV}) =-(261 \pm 6 \pm 8) $, where the first error is statistical 
and the second is systematic. Our results  are thus very compatible within uncertainties.
Note, however, that the above quoted lattice value \cite{SDlatt_recent} was 
obtained by fixing the scale with the kaon
decay constant $F_K$, determined in the quenched approximation.
Overall, recent $n_f=2$ lattice determinations of the condensate in the chiral limit 
from several independent methods 
are much more precise than those for $n_f=3$. We quote the 
estimate recently performed in \cite{LattFLAG19}, 
by combining results from \cite{SDlatt_recent,qqlatt_n2}: 
$|\la\bar q q\ra|^{1/3}_{n_f=2} = -(266\pm 10)$ MeV, where the uncertainties include both systematic 
and statistical ones.

One may also compare with recent results from spectral sum 
rules~\cite{qqSRlast}: $\la\bar u u\ra^{1/3}(\equiv \la\bar d d\ra^{1/3})(\mbox{2 GeV})\simeq -(276\pm 7)$ MeV. But
keeping in mind that 
the latter sum rules actually determine precisely the current quark masses, so that 
the $\la\bar u u\ra$ value is indirectly extracted from using the GMOR relation~(\ref{GMOR}). Accordingly
the comparison is not strictly for the chiral limit. In this context,
even though the overall reliability of $\bar\Lam(2)$ is not yet at the level of  $\bar\Lam(3)$,
the results (\ref{finn2}) and (\ref{qqfin2}) constitute reasonably accurate independent determinations
in the chiral limit. 
Indeed, given that the present $n_f=2$ lattice results for the condensate are quite accurate, 
it is tempting alternatively to combine the latter with our firmer result Eq.(\ref{finn2}), 
in order to rather determine a new 
independent estimate of $\bar\Lam(2)$: taking the above quoted estimate of the condensate given by \cite{LattFLAG19}, 
this gives 
\be
\bar\Lam(n_f=2) = (308 ^{+4}_{-6} \pm 12 ) \mbox{MeV} .
\ee
\subsection{ $n_f=0$ (quenched approximation)}
Finally for completeness we also give results for the quenched approximation ($n_f=0$).
In this case we evolve $\alpha_S(\mu)$ and use Eq.~(\ref{qqRGI})
at five loops but in the appropriate $n_f=0$ approximation. 
As previously we need
$\bar\Lam(n_f=0)$, available from various different approaches with lattice simulations.
We rely on the average performed in \cite{LattFLAG19}, combining different precise  
lattice results\cite{lattLam0}:
\be
\bar\Lam(n_f=0) =(257 \pm 7)\, \mbox{MeV} .
\label{Lam0}
\ee
As stressed in \cite{LattFLAG19}, it is worth noting that this value is obtained by using the
same value as for $n_f=2$ and $n_f=2+1$ of the basic lattice scale, defined from the quark static potential, 
$r_0=0.472\mbox{fm}$, which 
for $n_f=0$ amounts merely to a defining convention for $\bar\Lam(n_f=0)$. \\
Next the RG evolution from Eq.~(\ref{qqfin0}) leads to
\be
 \la\bar q q\ra^{1/3}_{n_f=0}(2 {\rm GeV}) =  -(0.932_{-.018}^{+.022}) \bar\Lam_0 .
 \label{qqn02GeV}
\ee
Combining (\ref{Lam0}) with 
Eq.~(\ref{qqn02GeV}) we obtain
\be
\la\bar q q\ra^{1/3}_{n_f=0}(2 {\rm GeV},\rm \bar\Lam_0)=  -(240^{+6}_{-5} \pm 6) \,{\rm MeV}\,. 
\label{qqfin0}
\ee
It appears to us not easy to compare (\ref{qqfin0}) with other determinations, since most phenomenological
determinations of the condensate are obviously obtained for $n_f\ge 2$.  
Concerning lattice simulations, 
most of the modern calculations no longer use the quenched approximation, performing simulations with  
fully dynamical sea quarks, while too old results in the quenched approximation are presumably affected 
by rather large uncertainties. 
To our knowledge, there is one precise, often quoted latest quenched simulation result\cite{qqlattn0}:
\be
\la\bar q q\ra^{1/3}_{n_f=0}(2 {\rm GeV},lattice)= -(250 \pm 3) \,{\rm MeV}\,.
\ee
So our result (\ref{qqfin0}) appears consistent with the latter within uncertainties.
We stress, however, that our study of the quenched case $n_f=0$ is merely motivated as a 
consistency crosscheck of our method, since the quenched approximation is anyway not very realistic. 
\subsection{Further discussion on $\la \bar q q \ra (n_f)$ dependence}
Comparing all our results for $n_f=0, 2, 3$ at the same perturbative orders, it appears that the {\em ratio} of 
the quark condensate to $\bar\Lam^3$ has a sizable but moderate dependence 
on the number of flavors $n_f$: there is a clear trend
that $|\la\bar q q\ra^{1/3}|(n_f)/\bar\Lam(n_f) $ decreases regularly, roughly 
linearly by about $4\%$ for $n_f \to n_f+1 $ (that is clear at least from the studied $n_f \le 3$ cases). 
Naively (perturbatively) the moderate dependence on $n_f$
is expected (as long as $n_f$ is not large), since it only appears
explicitly at three-loop order. Nevertheless it does not imply a similar decrease of the
absolute condensate values, as those depend on $\bar\Lam(n_f)$,
that appears rather to increase with $n_f$ for $n_f\le 3$ (at least if considering the
low lattice $\bar\Lam(0)$ value (\ref{Lam0}, but it is not so clear from comparing $\bar\Lam(2)$
and $\bar\Lam(3)$ given all the present uncertainties in their values).
Concerning the $n_f=3$ to $n_f=2$ condensate ratio, various 
lattice results have still rather large uncertainties at present~\cite{LattFLAG19} 
but some recent results are more compatible with a ratio unity\cite{ss_uulattice,Lattqqn3_corr}. 
The spectral sum rules prediction for the ratio is also not very precise~\cite{ss_uuSR,ss_uuSRlast}:
$\la\bar s s\ra/ \la\bar u u\ra = 0.74^{+0.34}_{-0.12}$, (see also the recent review \cite{qqSRrev18}).
Since our results are by construction valid in the strict chiral limit, taken at face value 
they indicate that the possibly larger difference 
obtained by some other determinations~\cite{LattFLAG19,qqflav} is more likely due to the
explicit breaking from the large strange quark mass, rather than an intrinsically strong $n_f$ dependence 
of the condensate in the exact chiral limit. 
\section{Summary and Conclusion}
We have reconsidered our variational RGOPT approach applied to the spectral density of the Dirac 
operator, the latter being obtained in a first stage from the perturbative logarithmic discontinuities of 
the quark condensate in the $\ms$ scheme.
This construction allows successive sequences of nontrivial variationally optimized results in the 
strict chiral limit, from 
two- to five-loop levels using exactly known perturbative content, and partially up to six loops, 
the latter more approximately relying on the six-loop content
exactly predictable from five-loop renormalization group properties.
The results Eqs.~(\ref{finn0})-(\ref{finn3}) 
are those that we consider the firmer, 
while latter results in Eqs.~(\ref{qqfin2}),(\ref{qqfin3wa}) are further affected by present 
uncertainties in perturbative evolution and $\bar\Lam$ values. 
 Eqs.~(\ref{finn0})-(\ref{finn3}) show a very good stability and empirical convergence, 
although the strictly five-loop results exhibit some
instabilities with respect to both four- and six-loop results. Those instabilities are traced to specific features of
the spectral density, namely the occurence at growing orders of new large $\pi^{2k}$ discontinuity contributions
that tend to destabilize the original perturbative coefficients when first appearing at a given order.
For all the considered cases from $n_f=0$ (quenched approximation)  to $n_f=3$,
it is striking that the six-loop results are very close to the four-loop ones, both exhibiting very stable
properties. It appears convincing to us that the systematically $\sim 2\%$ higher values of 
$|\la\bar q q\ra|^{1/3}_{RGI}(n_f)$ obtained at five-loop RGOPT order are largely 
an artifact of the instability from the $\pi^4$ discontinuity terms appearing first at five-loop order.
Nevertheless we  incorporate  more conservatively 
the differences between four-, five- and six-loop results as intrinsic theoretical uncertainties, 
which are of order $\pm 2\%$. Notice that, if less conservatively discarding the presumably less reliable 
strictly five-loop results from our averages, the lowest values in Eqs.~(\ref{finn0})-(\ref{finn3}) 
are favored, with much smaller uncertainties.
In any case the final condensate values and uncertainties in Eqs.(\ref{qqfin3wa}), (\ref{qqfin2}) are 
more affected by the present uncertainties on the basic QCD scale $\bar\Lam$, both for $n_f=2$ and $n_f=3$. 
(To possibly get rid of $\bar\Lam$ uncertainties, particularly for $n_f=2$, 
one could in principle apply RGOPT directly to a more physical
RG invariant quantity, like $ \la\bar q q\ra_{RGI}/F^3_\pi$, combining the present analysis
with the one in \cite{rgopt_alphas} for $F_\pi$: but this involves somewhat nontrivial issues 
and is left for future investigation).

In conclusion the chiral condensate values obtained in our analysis are very compatible, within uncertainties, 
with the most precise recent lattice determinations for all considered $n_f$ values. 
Our results for $n_f=3$ are perhaps of particular interest, given that other independent determinations 
are either not in the chiral limit or, concerning lattice results, are affected by still  
rather important uncertainties in the chiral extrapolation\cite{LattFLAG19}.  
Finally our results indicate a moderate flavor dependence of 
the $\la\bar q q\ra^{1/3}_{n_f}/\bar\Lam_{n_f}$ values in the chiral limit for $n_f \le 3$.
\acknowledgments
We are very grateful to Konstantin Chetyrkin for valuable exchanges, and to him and Pavel Baikov for providing us 
their results of the five-loop vacuum anomalous dimension~\cite{ga05l} before it was published.
We are also very grateful to Andreas Maier and Peter Marquard for valuable discussions at the RADCOR 2019 
conference, and for providing us
four-loop results\cite{maier0919}  
leading to Eq.(\ref{c44ex}) in our normalization. 
\appendix
\section{RG and other perturbative quantities}
In this appendix we give for completeness all the relevant quantities related to perturbative RG properties
used in our calculations.
The RG coefficients up to five loops for a general gauge theory were obtained in complete analytical form 
respectively in Refs.\cite{beta5l} for the beta function and \cite{gam5l} for the anomalous mass dimension. 
We do not repeat those expressions explicitly here, referring to these articles. 
Note simply that we mainly use the normalization $g\equiv 4\pi \alpha_S(\mu) \equiv 4\pi^2 a_s(\mu)$, such that 
\be
\beta(g)\equiv \frac{{\rm d}g}{ {\rm d}\ln\mu } = -2\sum_k b_k g^{k+2} , \;\; 
\gamma_m(g) = \sum_k \gamma_k g^{k+1} ,
\ee
where our  $b_i$, $\gamma_i$ expressions are related as $b_k = \beta_k (4\pi^2)^{-1-k}$, 
$ \gamma_k =2 (4\pi^2)^{-1-k}\gamma_k$ with respect {\em e.g.} to 
the first refs. in \cite{beta5l}, \cite{gam5l} respectively. 
\subsection{Vacuum energy anomalous dimension}
Next, in our normalization conventions the anomalous dimension of the vacuum energy, $\Gamma^0(a_s)$ entering Eq.(\ref{RGanom}), is 
given to five-loop order as
\be
\Gamma^0(a_s) = -2 \times (\frac{3}{16\pi^2})
\left[ 1 +  \sum_{k=1}^4 a_s \Gamma^0_k +{\cal O}(a_s^5) \right] 
\label{Gam05l}
\ee
with the coefficients up to three loops determined long ago~\cite{vac_anom3}:
\be
\Gamma^0_1 =\frac{4}{3} , \;\;
\Gamma^0_2 =\frac{457}{72} -\frac{5}{12} n_l -\frac{29}{12} n_h 
-\frac{2}{3} z_3 ,
\label{gam03l}
\ee
and the four-loop $\Gamma^0_3$ and five-loop $\Gamma^0_4$ coefficients, obtained in full analytical
form in ref.\cite{ga05l}, are given explicitly 
respectively in Eqs. (3.4), (3.5) of \cite{ga05l}, for $n_f\equiv n_l+n_h$ quark flavors
and $n_h=1$ heavy quark\footnote{Note a trivial factor 2 normalization difference in Eq.(\ref{Gam05l}) with respect to 
\cite{ga05l} due to our use of $d\ln\mu $ in Eq.(\ref{RGanom}). Also $\Gamma^0_k$ in 
Eqs.(\ref{Gam05l})-(\ref{gam05l}) 
differ by an overall factor $3/(16\pi^2)$ from the original
$(\gamma_0^{di,nd})_k$ in the notations of \cite{ga05l}.}. 
For completeness we give here their relevant expressions adapted to our case
(in numerical form for short), where in practice we consider $n_l=0$ and $n_h (\equiv n_f)$
{\em massive} degenerate quarks:
\bea
\Gamma^0_3(n_l,n_h)\simeq &&
 33.6625 + 0.18139 n_h^2  - 32.5586 n_h \nn \\ 
&& + 0.214632 n_h n_l - 4.96507 n_l +  0.0332417 n_l^2 ,
\label{gam04l}
\eea
and 
\bea
 \Gamma^0_4(n_l,n_h)\simeq &&
242.021 + 0.0185562 n_h^3 + 17.6037 n_h^2 + 0.03715 n_h^2 n_l \nn \\
&& -43.3192 n_l +  1.0631 n_l^2 + 0.0000376492 n_l^3   \nn \\
&& -299.998 n_h + 18.6668 n_h n_l + 0.0186315 n_h n_l^2 .
\label{gam05l}
\eea
\subsection{Perturbative condensate and spectral density}
Next, the coefficients
of the perturbative quark condensate,  Eq.(\ref{qqQCDpert}), were given at three loops in Eq.(\ref{c3i}). 
Using Eqs.(\ref{RGanom}) with Eq.(\ref{RGop}) we determine the relevant coefficients
at four-loop and higher orders:
\bea
& c_{40}& = \frac{1}{54} (81-2n_f)(57-2 n_f) \nn \\
& c_{41}& =-\frac{1}{81} (18171 +20 n_f^2 - 1353 n_f )  \nn \\
& c_{42}& =\frac{1}{8\times 81} \left[ 226647-30690 n_h -18378 n_l +632n_h^2 +832 n_h n_l +200 n_l^2 -72(57 +28 n_f) z_3
\right]  \nn \\
& c_{43}& = \frac{1}{32\times 729} \left[ 4 n_h^2 (4752 z_3-5935) -4 n_l^2 (3024 z_3+751) +8 n_l n_h (864 z_3-3343) 
\right. \nn \\
&& \left. -12 n_f ( 2304 a_4+96 l_{z2}-5040z_4 ) 
+12n_l (22428 z_3 +53093) -12 n_h (30060 z_3+9720 z_5-124373) \right. \nn \\ 
&& \left. +9 (87552 a_4+3648 l_{z2}+10368 z_3-68400 z_4-81840z_5-854633)  \right]
\label{c4k}
\eea
where $n_f=n_l +n_h$, $l_{z2}= \ln^2(2)(\ln^2(2)-6z_2)$, $z_i =\zeta(i)$.\\
In numerical approximation this gives
\bea
& c_{40}& = 85.5 - 5.11111 n_f + 0.0740741 n_f^2 \nn \\
& c_{41}& = -224.333 + 16.7037 n_f - 0.246914 n_f^2 \nn \\
& c_{42}& = 342.151  -51.1008 n_h - 32.1008 n_l + 0.975309 n_h^2 + 0.308642 n_l^2 + 
+ 1.28395 n_h n_l  \nn \\
& c_{43}& = -375.082 + 42.6214 n_h + 43.5949 n_l - 0.0382074 n_h^2  - 
 0.790268 n_h n_l - 0.752061 n_l^2 \,,
\eea
while the complete expression for the last nonlogarithmic four-loop coefficient $c_{44}$  
is given explicitly in Eq.(\ref{c44ex}).\\
Similarly we obtain for the five-loop logarithmic coefficients:
\be
c_{50}  = -\frac{1}{4\times 135} (81-2n_f)(57-2n_f)(49-2n_f) 
\label{c50}
\ee
\be
c_{51}  = -\frac{1}{8\times 81} (-952101 +40 n_f^3 -3898 n_f^2 +112749 n_f ) 
\label{c51}
\ee
\bea
& c_{52} & = \frac{1}{16\times 243}  \left( 2150625 n_h -95963 n_h^2 +1264 n_h^3 -146134 n_h n_l \right. \nn \\ 
&& \left. +2928 n_h^2 n_l-50171 n_l^2 +2064 n_h n_l^2 +400 n_l^3 \right. \nn \\ 
&& \left. -144 n_f (-869+28n_f ) z_3 + 3(515779 n_l +72 (-56554+931 z_3)) \right) 
\label{c52}
\eea
\bea
& 729 \times 4^3\, c_{53} & =  32 n_h^3 (648 z_3-773) -32 n_l^3 (324 z_3+125) \nn \\
&& +9 \left ( -(768 a_4 +32 l_{z2})(57-2n_f)(49-2n_f) \right. \nn \\
&& \left. +116448 z_3+1675800  z_4+1292280 z_5+23864201\right )  \nn \\
&& +n_h (-4320 (27  n_f-836) z_5+6310872  z_3-1455840 z_4-50009817)
  \nn \\ 
&& +96 n_l n_h^2 (324 z_3-557) +n_l n_h (-131328 z_3+43200 z_4+3132538) \nn \\
&& -15 n_l (608040 z_3+97056 z_4-50256 z_5+1871063) \nn \\ 
&&   +n_h^2 (-761616 z_3+21600 z_4+2316653) \nn \\
&& +n_l^2 (-32736 n_h +630288 z_3+21600 z_4+815885) 
\label{c53}
\eea
\bea \displaystyle
& 243\times 2\times 4^3 \; c_{54} & = - 243\times 4^3\, (49-2n_f) c_{44} \nn \\
&& +n_h^3 (48 z_3-1728 z_4+875) 
 +n_l^3 (48 z_3-1728 z_4+1451) \nn \\
&&   +n_h^2 n_l \left( 144 z_3-5184 z_4+3201 \right) \nn \\
 &&  +n_h n_l^2 \left( 144 z_3-5184 z_4+3777 \right) \nn \\
&& +n_h n_l \left(-144 z_3   (81 z_3+2735)+128736 z_4+268992 z_5-48600 z_6
 -\frac{9903989}{18}\right)  \nn \\ 
&& +n_h^2 \left( -72 z_3 (162
   z_3+3191)+96444 z_4+173376
   z_5-48600
   z_6-\frac{16643189}{36}\right) \nn \\
&&  +n_l^2
   \left(-164088 z_3+32292 z_4+95616
   z_5-\frac{3164789}{36}\right) \nn \\
&& +n_h \left( -162816 a_4-6784 l_{z2}  +12 z_3
   (39228 z_3+407969)-932766 z_4 \right. \nn \\ 
   && \left.   -8087688
   z_5+262800 z_6+3213000
   z_7+\frac{54045443}{6} \right) \nn \\ 
&&   +n_l  \left( -162816 a_4-6784 l_{z2}+12
   (134429-12990 z_3) z_3     \right. \nn \\ 
&& \left.   +638958
   z_4-461448 z_5-927900 z_6-95256
   z_7+\frac{21934883}{6} \right) \nn \\
   &&   +1170432
   a_4+48768 l_{z2}
        +12 z_3
   (8031 z_3+1007870)-5759550 z_4 \nn \\
&&   -27852768 z_5+10192050
   z_6+8641836 z_7-\frac{145813179}{8} \; ,
\label{c54}
\eea
where we conveniently expressed $c_{54}$
in terms of the four-loop nonlogarithmic $c_{44}$ coefficient.
In numerical approximation we obtain: 
\bea
& c_{50}& = -418.95 + 42.1444 n_f - 1.38519 n_f^2 + 0.0148148 n_f^3 \nn \\
& c_{51}& = 1469.29 - 173.995 n_f + 6.01543 n_f^2 - 0.0617284 n_f^3 \nn \\
& c_{52}& =  -3079.72 + 436.666 n_f - 14.1506 n_f^2 + 0.102881 n_f^3 \nn \\
&& + n_h (155.167 - 
 11.7778 n_f  + 0.222222 n_f^2) \nn \\
& c_{53}& = 5102.45 - 852.446 n_h - 843.205 n_l + 61.7769 n_h n_l \nn \\
&& + 27.7449 n_h^2 + 34.032 n_l^2
- 0.344719 n_h^2 n_l \nn \\
&& - 0.701646 n_h n_l^2 
 + 0.00406919 n_h^3  -  0.352858 n_l^3 \nn \\
& c_{54}& =  (n_f- 24.50) c_{44}
 -617.146  + 309.613 n_h + 144.324 n_l \nn \\ 
&& - 16.7381 n_h^2 - 4.8565 n_l^2
  - 21.5946 n_h n_l \nn \\
&& - 0.0719093 n_h^2 n_l  
- 0.0533908 n_h n_l^2 
- 0.0301426 n_h^3  - 0.0116241 n_l^3
\eea
It is straightforward to apply the RG Eq.(\ref{RGop}) to obtain similarly the six-loop logarithmic
coefficients $c_{6k}$ of $L_m^{6-k}$. To avoid unnecessarily lengthy expressions it is convenient to equivalently 
express the $c_{6k}$ as functions of the above lower order perturbative and RG coefficients: with the same
normalization as in Eq.(\ref{qqQCDpert}), with an overall $3/(2\pi^2) m^3 a_s^5$ factor at six-loops, they read:
\bea \displaystyle
& c_{60} = -\frac{8\pi^2}{3}\,(2 b_0 + \gamma_0) ,\nn \\
& c_{61} = -\frac{4\pi^2}{5} \, \left(4 (2 b_0 +\gamma_0) c_{51} + 5 c_{50} \gamma_0 + 
   8 \pi^2\,c_{40} (3 b_1 + 2 \gamma_1) \right) \nn \\
& c_{62} = -4 \pi^2 \left[ (2 b_0+\gamma_0) c_{52} + \gamma_0 c_{51}  
- 2\pi^2 \big( (3 b_1 +2\gamma_1) c_{41} + 2 c_{40} \gamma_1 \big) \right. \nn \\ 
& \left. - 16\pi^4 c_{30} (b_2 + \gamma_2)  \right] \nn \\
& c_{63} = -\frac{4}{3} \pi^2\,\left[ 3 \gamma_0 c_{52} + 4 (2 b_0 + \gamma_0) c_{53} + 
   4 \pi^2 \left( 2(3 b_1 +2 \gamma_1) c_{42} + 3 \gamma_1 c_{41} \right.\right. \nn \\
&  \left.\left.  + 
      4 \pi^2 (3 c_{30} \gamma_2 + 4 c_{31} (b_2 + \gamma_2) + 
         8 \pi^2 c_{20} (b_3 + 2 \gamma_3) \,) \; \right) \;\right] , \nn \\
& c_{64} = 4 \pi^2\, \left[ -2 (2 b_0+\gamma_0) c_{54} -\gamma_0 c_{53} 
-4 \pi^2 \left( (3 b_1+2\gamma_1) c_{43} 
+\gamma_1 c_{42} \right) \right. \nn \\
& \left. -16 \pi^4 \left(2 c_{32} (b_2+\gamma_2) +c_{31} \gamma_2 \right) \right. \nn \\ 
& \left. -64 \pi^6 \left( (b_3 +2\gamma_3) c_{21} +\gamma_3 c_{20} \right) 
 +512 \pi^8 \gamma_4 \right]  .
\label{c6k}
\eea
Next from Eq.~(\ref{discgen}), (\ref{disc3}) it is straightforward to derive the corresponding six-loop
coefficients of the spectral density in the normalization of Eq.~(\ref{SD5l}), that we give here for completeness:
\bea
&\rho_{61} & = 3 c_{60} ,\nn \\
&\rho_{62} & = \frac{5}{2} c_{61} ,\nn \\
&\rho_{63} & = 2 c_{62} -\frac{5\pi^2}{2} c_{60} ,\nn \\
&\rho_{64} & = \frac{3}{2} c_{63} -\frac{5\pi^2}{4} c_{61},\nn \\
&\rho_{65} & = c_{64} -\frac{\pi^2}{2} c_{62} +\frac{3\pi^4}{16} c_{60} ,\nn \\
&\rho_{66} & =  \frac{1}{2} c_{65} -\frac{\pi^2}{8} c_{63} +\frac{\pi^4}{32} c_{61} .
\label{rho6k}
\eea
Note that from standard RG properties, only $c_{65}$ (and therefore only the nonlogarithmic coefficient 
$\rho_{66}$ of $\rho(\lambda)$ in Eq.(\ref{rho6k})) depend on
the presently unknown six-loop vacuum energy $\Gamma^0_5$ and nonlogarithmic five-loop coefficient $c_{55}$. 
Accordingly, as explained in the main text,
we simply ignore $c_{65}$ and $\rho_{66}$ in our six-loop analysis.
\subsection{RG invariant perturbative subtraction}
Next, we also derive for completeness the coefficients entering the subtraction function $S(m,g)$ defined in 
Eqs.(\ref{qq_sub}), (\ref{sub}),  
such that $m \la \bar q q \ra -S(m,g) $ defines a (finite) condensate obeying the homogenous RG equation
Eq.(\ref{RGop}) up to five loops, that can be useful for different purposes. 
We obtain after some algebra\footnote{The normalization of $s_i$ coefficients in Eq.(\ref{sub}) is different
from the one in ref.\cite{rgoqq1} for convenience, but Eq.(\ref{s_i}) is consistent with our previous expressions
up to three-loop order in \cite{rgoqq1}.}  
\bea
&& s_0 = \frac{1}{8\pi^2 (b_0-2\, \gamma_0)} = -\frac{6}{ (15+2n_f)}\, , \nn \\
&& s_1= (2\ga_0)^{-1} \left(  4\pi^2 (b_1-2\, \gamma_1)\, s_0 -\frac{\Gamma^0_1}{8\pi^2}  \right) 
=  \frac{633+n_f}{48(15+2n_f)} , \nn \\
&& s_2 = \left(2 \ga_0 +b_0\right)^{-1}\left( (4\pi^2)^2 (b_2-2 \ga_2)\,s_0
 -2 (4\pi^2) \ga_1\,s_1 -\frac{\Gamma^0_2}{8\pi^2} \right) \nn \\
&& =
 \frac{ n_h (-24519 + 33408 z_3) + n_l (1401 + 33408 z_3) 
-746 n_l n_h   -2101 n_h^2 + 1355 n_l^2 - 27 (4151 + 320 z_3) }
{144 \,(15 + 2 nf)(-81 + 2\, n_f) }\, , \nn \\
&& s_3 = 
 \left(2\ga_0 +2b_0\right)^{-1}\left( (4\pi^2)^3 (b_3-2\, \ga_3)\, s_0
 -2 (4\pi^2)^2 \ga_2 \, s_1 -4\pi^2(2\ga_1+b_1) s_2 -\frac{\Gamma^0_3}{8\pi^2} \right) , \nn \\
&& s_4 =  \left(2 \ga_0 +3b_0 \right)^{-1}\left(
 (4\pi^2)^4 (b_4-2\ga_4)\,s_0  -2(4\pi^2)^3 \ga_3 s_1 -(4\pi^2)^2 (2\ga_2+b_2)\,s_2 -
2(4\pi^2) (\ga_1+b_1)\,s_3 -\frac{\Gamma^0_4}{8\pi^2}  \right)
\label{s_i}
\eea
where we give both generic compact expressions and their particular QCD values, 
%
%
%
%
%
%
%
%
%
%
%
%
the latter for the four-loop $s_3$ and five-loop $s_4$ coefficients 
given numerically to $10^{-6}$ accuracy as:
\bea
&s_3 = & \frac{32}{(-81 + 2 n_f) (-57 + 2 n_f) (15 + 2 n_f)} \times \nn \\
&& \left( 2485.78 + 1045.17 n_h  - 1314.62 n_l + 436.946 n_h n_l \right. \nn \\ 
&& \left. + 351.286 n_h^2 + 85.66 n_l^2 - 8.6872 n_h^3  - 2.56879 n_l^3 \right. \nn \\ 
&& \left. - 19.9432 n_h^2 n_l  - 13.8248 n_h n_l^2 + 0.00262828 n_h^2 n_l^2 \right. \nn \\ 
&& \left. + 0.0573077 n_h^3 n_l 
 - 0.0538034 n_h n_l^3 + 0.0282158 n_h^4 - 0.0273397 n_l^4 \right) ,
\label{s3num} 
\eea 

\bea
&s_4 = & \frac{16}{(-49 + 2 n_f)(-81 + 2 n_f) (-57 + 2 n_f) (15 + 2 n_f)} \times \nn \\
&& \left( -1947880. + 0.0102918 n_h^6 + 13.0258 n_h^5  + 0.0432323 n_h^5 n_l \right. \nn \\ 
&& \left. + 682641.\, n_l -  98179.9 n_l^2 + 3437.72 n_l^3 + 42.1262 n_l^4 \right. \nn \\ 
&& \left. - 1.59314 n_l^5 - 0.00822672 n_l^6  -943.625 n_h^4  + 50.5099 n_h^4 n_l \right. \nn \\ 
&& \left. + 0.0617844 n_h^4 n_l^2 + 19279.\, n_h^3  - 2788.75 n_h^3 n_l + 71.782 n_h^3 n_l^2 \right. \nn \\ 
&& \left. + 0.0206508 n_h^3 n_l^3  -105289.\, n_h^2  + 41995.6 n_h^2 n_l - 2704.5 n_h^2 n_l^2 \right. \nn \\ 
&& \left. + 42.5442 n_h^2 n_l^3 - 0.0308082 n_h^2 n_l^4  -723426. n_h  - 203469. n_h n_l \right. \nn \\ 
&& \left. + 26154.4 n_h n_l^2 -   817.246 n_h n_l^3 + 6.65319 n_h n_l^4 - 0.0308418 n_h n_l^5      
   \right) .
\eea


\begin{thebibliography}{999}

\bibitem{GMOR} M.~Gell-Mann, R.~J.~Oakes and B.~Renner,
   Phys.\ Rev.\  {\bf 175}, 2195 (1968).
%
\bibitem{LattFLAG19} 
S.~Aoki {\it et al.} [Flavour Lattice Averaging Group],
  arXiv:1902.08191 [hep-lat].
%
%
\bibitem{qqSR}  See {\it e.g.} H.~G.~Dosch and S.~Narison,
  Phys.\ Lett.\ B {\bf 417}, 173 (1998); 
   M. Jamin, 
   Phys.\ Lett.\ B {\bf 538}, 71 (2002).
%
\bibitem{qqSRlast} 
S.~Narison,
  Phys.\ Lett.\ B {\bf 738}, 346 (2014)
  [arXiv:1401.3689 [hep-ph]].
%
%
\bibitem{SVZSSR}
M.~A.~Shifman, A.~I.~Vainshtein, and V.~I.~Zakharov,  Nucl.\ Phys.\ {\bf B147}, 385 (1979).
%
\bibitem{NJL} Y.~Nambu and G.~Jona-Lasinio,
  Phys.\ Rev.\  {\bf 122}, 345 (1961).
%
\bibitem{NJLrev} S.~P.~Klevansky,
  Rev.\ Mod.\ Phys.\  {\bf 64}, 649 (1992);
  T.~Hatsuda and T.~Kunihiro,
  Phys.\ Rept.\  {\bf 247}, 221 (1994).
%
\bibitem{DSErev}
C.~D.~Roberts and S.~M.~Schmidt,
  Prog.\ Part.\ Nucl.\ Phys.\  {\bf 45}, S1 (2000);
R.~Alkofer and L.~von Smekal,
  Phys.\ Rept.\  {\bf 353}, 281 (2001).
%
  \bibitem{D-S} P.~Maris, C.~D.~Roberts, and P.~C.~Tandy, Phys.\ Lett.\ B {\bf 420}, 267 (1998);
 P.~Maris and C.~D.~Roberts, Phys.\ Rev.\ C {\bf 56}, 3369 (1997).
%
  \bibitem{qqbar_link} K.~Langfeld, H.~Markum, R.~Pullirsch, C.~D.~Roberts, and S.~M.~Schmidt,
  Phys.\ Rev.\ C {\bf 67}, 065206 (2003).
%
\bibitem{DSEvar} D.~R.~Campagnari, E.~Ebadati, H.~Reinhardt and P.~Vastag,
  Phys.\ Rev.\ D {\bf 94}, no. 7, 074027 (2016)
  [arXiv:1608.06820 [hep-ph]].
%
\bibitem{qqlatt_other} See {\it e.g.}  D.~Becirevic and V.~Lubicz,
  Phys.\ Lett.\ B {\bf 600}, 83 (2004);
 P.~Hernandez, K.~Jansen, L.~Lellouch, and H.~Wittig,
  J. High Energy Phys. {\bf 0107}, 018 (2001);
  A.~Duncan, S.~Pernice, and J.~Yoo,
  Phys.\ Rev.\ D {\bf 65}, 094509 (2002).
%
\bibitem{qqlattSDearly}
 E.~Marinari, G.~Parisi, and C.~Rebbi, Phys.\ Rev.\ Lett.\ {\bf 47}, 1795 (1981).
%
\bibitem{qqlattSD}
  L.~Del Debbio, L.~Giusti, M.~L\"uscher, R.~Petronzio, and N.~Tantalo,
  J. High Energy Phys. {\bf 0602}, 011 (2006);
 L.~Giusti, and M.~L\"uscher, J.\ High Energy Phys. 03 ({\bf 2009}) 013;
 H.~Fukaya, S.~Aoki, T.W.~Chiu, S.~Hashimoto,
T.~Kaneko, J.~Noaki, T.~Onogi, and N.~Yamada, Phys.\ Rev.\ Lett.\ {\bf 104}, 122002 (2010) [Erratum-{\it ibid.}\  {\bf 105}, 159901 (2010)].
%
\bibitem{SDlatt_recent} 
G.~P.~Engel, L.~Giusti, S.~Lottini and R.~Sommer,
  Phys.\ Rev.\ D {\bf 91}, no. 5, 054505 (2015)
  [arXiv:1411.6386 [hep-lat]].
%
\bibitem{BanksCasher}  T.~Banks and A.~Casher, Nucl.\ Phys.\ {\bf B169}, 103 (1980);
%
\bibitem{SDgen} H.~Leutwyler and A.~V.~Smilga,
  Phys.\ Rev.\ D {\bf 46} (1992) 5607.
%
\bibitem{SDchpt2}  
  A.~V.~Smilga and J.~Stern,
  Phys.\ Lett.\ B {\bf 318}, 531 (1993);
  K.~Zyablyuk,  J. High Energy Phys. {\bf 0006}, 025 (2000).
%
\bibitem{CSBlatt_rev18} See e.g. for a recent review, 
M.~Faber and R.~H\"ollwieser,
  Prog.\ Part.\ Nucl.\ Phys.\  {\bf 97}, 312 (2017)
  [arXiv:1908.09740 [hep-lat]].
%
\bibitem{chpt}
J.~Gasser and H.~Leutwyler,
  Annals\ Phys.\  {\bf 158}, 142 (1984);
Nucl.\ Phys.\ {\bf B250}, 465 (1985).
%
 \bibitem{chptn3} J.~Bijnens and T.~A.~Lahde,
  Phys.\ Rev.\ D {\bf 71} (2005) 094502;
  J.~Bijnens and G.~Ecker,
  Ann.\ Rev.\ Nucl.\ Part.\ Sci.\  {\bf 64}, 149 (2014)
  [arXiv:1405.6488 [hep-ph]].
%
  \bibitem{qqflav} 
S.~Descotes-Genon, L.~Girlanda, and J.~Stern, J.\ High\ Energy\ Phys.\ 01 {\bf 2000} 041;  
S.~Descotes-Genon and J.~Stern, Phys.\ Lett.\ B {\bf 488}, 274 (2000);
S.~Descotes-Genon, N.~H.~Fuchs, L.~Girlanda, and J.~Stern, Eur.\ Phys.\ J.\ C {\bf 34}, 201 (2004);
V.~Bernard, S.~Descotes-Genon, and G.~Toucas, J.\ High\ Energy\ Phys.\ 01 ({\bf 2011}) 107;
V.~Bernard, S.~Descotes-Genon, and G.~Toucas,
  J. High Energy Phys. {\bf 1206}, 051 (2012);
M.~Kolesar and J.~Novotny,
  Eur.\ Phys.\ J.\ C {\bf 78}, no. 3, 264 (2018)
  [arXiv:1709.08543 [hep-ph]].
%
\bibitem{rgopt1} J.-L.~Kneur and A.~Neveu, Phys.\ Rev.\ D {\bf 81}, 125012 (2010). 
%
\bibitem{rgopt_Lam}  J.-L.~Kneur and A.~Neveu,
  Phys.\ Rev.\ D {\bf 85}, 014005 (2012).
%
\bibitem{rgopt_alphas}  J.-L.~Kneur and A.~Neveu,
 Phys.\ Rev.\ D {\bf 88}, 074025 (2013).
%
  \bibitem{rgoqq1} J.~L.~Kneur and A.~Neveu,
  Phys.\ Rev.\ D {\bf 92}, no. 7, 074027 (2015)
  [arXiv:1506.07506 [hep-ph]].
\bibitem{rgopt_phi4}
J.-L.~Kneur and M.~B.~Pinto,
  Phys.\ Rev.\ Lett.\  {\bf 116}, no. 3, 031601 (2016)
  [arXiv:1507.03508 [hep-ph]];
  Phys.\ Rev.\ D {\bf 92}, no. 11, 116008 (2015)
  [arXiv:1508.02610 [hep-ph]].
%
\bibitem{rgopt_nlsm}
G.~N.~Ferrari, J.~L.~Kneur, M.~B.~Pinto and R.~O.~Ramos,
  Phys.\ Rev.\ D {\bf 96}, no. 11, 116009 (2017)
  [arXiv:1709.03457 [hep-ph]].
%
\bibitem{rgopt_qqmu}
J.~L.~Kneur, M.~B.~Pinto and T.~E.~Restrepo,
  Phys.\ Rev.\ D {\bf 100}, no. 11, 114006 (2019)
  [arXiv:1908.08363 [hep-ph]].
%
\bibitem{beta5l} P.~A.~Baikov, K.~G.~Chetyrkin and J.~H.~K\"uhn,
  Phys.\ Rev.\ Lett.\  {\bf 118}, no. 8, 082002 (2017);
  [arXiv:1606.08659 [hep-ph]];
  T.~Luthe, A.~Maier, P.~Marquard and Y.~Schr\"oder,
  JHEP {\bf 1607}, 127 (2016);
  [arXiv:1606.08662 [hep-ph]];
  F.~Herzog, B.~Ruijl, T.~Ueda, J.~A.~M.~Vermaseren and A.~Vogt,
  JHEP {\bf 1702}, 090 (2017)
  [arXiv:1701.01404 [hep-ph]];
  T.~Luthe, A.~Maier, P.~Marquard and Y.~Schr\"oder,
  JHEP {\bf 1710}, 166 (2017); 
  [arXiv:1709.07718 [hep-ph]];
  K.~G.~Chetyrkin, G.~Falcioni, F.~Herzog and J.~A.~M.~Vermaseren,
  JHEP {\bf 1710}, 179 (2017)
  Addendum: [JHEP {\bf 1712}, 006 (2017)]
  [arXiv:1709.08541 [hep-ph]].
  
\bibitem{gam5l} P.~A.~Baikov, K.~G.~Chetyrkin and J.~H.~K\"uhn,
  JHEP {\bf 1410}, 076 (2014);
  [arXiv:1402.6611 [hep-ph]];
  T.~Luthe, A.~Maier, P.~Marquard and Y.~Schr\"oder,
  JHEP {\bf 1701}, 081 (2017);
  [arXiv:1612.05512 [hep-ph]];
%
\bibitem{ga05l}
P.~A.~Baikov and K.~G.~Chetyrkin,
  PoS RADCOR {\bf 2017}, 025 (2018).
%

\bibitem{delta} 
V.I. Yukalov, Theor. Math. Phys. {\bf 28}, 652 (1976);
W.E. Caswell, Ann. Phys. (N.Y) {\bf 123}, 153 (1979);
I.G.  Halliday and P. Suranyi, Phys. Lett. B
 {\bf 85}, 421 (1979); 
 P. M. Stevenson, Phys. Rev. D {\bf 23}, 2916 (1981);
Nucl. Phys. {\bf B203}, 472 (1982);
J. Killinbeck, J. Phys. A {\bf 14}, 1005 (1981);
R.P. Feynman and H. Kleinert, Phys. Rev. A {\bf 34}, 5080 (1986);
A. Okopinska, Phys. Rev. D {\bf 35}, 1835 (1987);
A. Duncan and M. Moshe, Phys. Lett. B {\bf 215}, 352 (1988);
H.F. Jones and M. Moshe, Phys. Lett. B {\bf 234}, 492 (1990);
A. Neveu, Nucl. Phys. B, Proc. Suppl. {\bf B18}, 242 (1991);
V. Yukalov, J. Math. Phys (N.Y.) {\bf 32}, 1235 (1991);
S. Gandhi, H.F. Jones and M. Pinto, Nucl. Phys. {\bf B359}, 429 (1991);
C.~M. Bender {\it et al.}, Phys. Rev. D {\bf 45}, 1248 (1992);
S. Gandhi and M.B. Pinto, Phys. Rev. D {\bf 46}, 2570 (1992);
H. Yamada, Z. Phys. C {\bf 59}, 67 (1993);
K.G. Klimenko, Z. Phys. C {\bf 60}, 677 (1993);
A.N. Sissakian, I.L. Solovtsov, and O.P. Solovtsova, 
Phys. Lett. B {\bf 321}, 381 (1994);
H. Kleinert, Phys. Rev. D {\bf 57}, 2264 (1998); Phys. Lett. B {\bf 434}, 74
(1998);  Phys. Rev. D {\bf 60}, 085001 (1999); Mod. Phys. Lett. B {\bf 17}, 1011 (2003). 
%
%
\bibitem{gn2} C.~Arvanitis, F.~Geniet, M.~Iacomi, J.-L.~Kneur, and A.~Neveu,
Int.\ J.\ Mod.\ Phys.\ A {\bf 12}, 3307 (1997).
%
\bibitem{qcd1} C.~Arvanitis, F.~Geniet, J.~L.~Kneur, and A.~Neveu,
  Phys.\ Lett.\ B {\bf 390}, 385 (1997).
%
\bibitem{qcd2}  J.-L. Kneur, Phys.\ Rev.\ D {\bf 57}, 2785 (1998).
%
%
\bibitem{odm} R. Seznec and J. Zinn-Justin, J. Math. 
 Phys. {\bf 20}, 1398 (1979); J.C. Le Guillou and J. Zinn-Justin,
Ann. Phys. {\bf 147}, 57 (1983); J. Zinn-Justin, arXiv:1001.0675.
%
\bibitem{deltaconv} R.~Guida, K.~Konishi, and H.~Suzuki, Ann. Phys. (N.Y.) {\bf 241}, 152
(1995); {\bf 249}, 109 (1996). 
%
\bibitem{beccrit} H.~Kleinert, Mod.\ Phys.\ Lett.\ B {\bf 17}, 1011 (2003);
 B.~Hamprecht and H.~Kleinert,
  Phys.\ Rev.\ D {\bf 68}, 065001 (2003);
 B.~Kastening, Phys.\ Rev.\ A {\bf 68}, 061601 (2003); Phys.Rev. A {\bf 69}, 043613 (2004).
%
\bibitem{bec2} J.-L.~Kneur, A.~Neveu, and M.~B.~Pinto,
Phys.\ Rev.\ A {\bf 69}, 053624 (2004).
%
\bibitem{vac_anom2} K.~G.~Chetyrkin and V.~P.~Spiridonov, Sov.\ J.\ Nucl.\ Phys.\ {\bf 47}, 522 (1988); 
%
\bibitem{vac_anom3}
K.~G.~Chetyrkin and J.~H.~K\"uhn, Nucl.\ Phys.\ {\bf B432}, 337 (1994);
K.~G.~Chetyrkin and A.~Maier,  J.\ High\ Energy\ Phys.\ 01 ({\bf 2010}) 092.
%
\bibitem{qq3l} K.~G.~Chetyrkin and A.~Maier (private communication, 2010).
%
\bibitem{bgam4loop} 
  J.~A.~M.~Vermaseren, S.~A.~Larin, and T.~van Ritbergen,
  Phys.\ Lett.\ B {\bf 405}, 327 (1997).
K.~G.~Chetyrkin, Nucl.\ Phys.\ {\bf B710}, 499 (2005);
M.~Czakon Nucl.\ Phys.\ {\bf B710}, 485 (2005).
%
\bibitem{maier0919} A.~Maier and P.~Marquard (private communication, Sept. 2019).
%
\bibitem{correl3lns}
K.~G.~Chetyrkin, J.~H.~Kuhn and M.~Steinhauser,
  Nucl.\ Phys.\ B {\bf 505}, 40 (1997);
  [hep-ph/9705254].
  \bibitem{correl3ls}
  K.~G.~Chetyrkin, R.~Harlander and M.~Steinhauser,
  Phys.\ Rev.\ D {\bf 58}, 014012 (1998)
  [hep-ph/9801432];
A.~Maier, P.~Maierhofer and P.~Marquard,
  Nucl.\ Phys.\ B {\bf 797}, 218 (2008)
  [arXiv:0711.2636 [hep-ph]].

\bibitem{correl4l} K.~G.~Chetyrkin, J.~H.~Kuhn and C.~Sturm,
  Eur.\ Phys.\ J.\ C {\bf 48}, 107 (2006)
  [hep-ph/0604234].
%
\bibitem{PiSns4l} C.~Sturm,
  JHEP {\bf 0809}, 075 (2008)
  doi:10.1088/1126-6708/2008/09/075
  [arXiv:0805.3358 [hep-ph]].
%
\bibitem{rho_latt_alphas}
K.~Nakayama, H.~Fukaya and S.~Hashimoto,
  Phys.\ Rev.\ D {\bf 98}, no. 1, 014501 (2018)
  [arXiv:1804.06695 [hep-lat]].
%

\bibitem{PDG} See the QCD review chapter by J. Huston, K. Rabbertz and G. Zanderighi in
M.~Tanabashi {\it et al.} [Particle Data Group],
  Phys.\ Rev.\ D {\bf 98}, no. 3, 030001 (2018).
%
\bibitem{GN} D.~J.~Gross and A.~Neveu, Phys. Rev. D {\bf 10}, 3235 (1974).
%
%
\bibitem{matching4l} K.G.~Chetyrkin, B.A.~Kniehl and M.~Steinhauser, Phys.\ Rev.\ Lett.\  {\bf 79} (1997) 2184
  [hep-ph/9706430]; 
  Nucl.\ Phys.\ {\bf B510}, 61 (1998); 
K.~G.~Chetyrkin, J.H K\"uhn and C.~Sturm,  Nucl.\ Phys.\ {\bf B744}, 121 (2006).
%
\bibitem{QCDruncode}
K.~G.~Chetyrkin, J.~H.~Kuhn and M.~Steinhauser,
  Comput.\ Phys.\ Commun.\  {\bf 133}, 43 (2000);
  F.~Herren and M.~Steinhauser,
  Comput.\ Phys.\ Commun.\  {\bf 224} (2018) 333
  [arXiv:1703.03751 [hep-ph]].
%
\bibitem{Lattqqn3} 
A.~Bazavov {\it et al.} [MILC Collaboration],
  PoS CD {\bf 09}, 007 (2009)
  [arXiv:0910.2966 [hep-ph]].
%
\bibitem{Lattqqn3_low}
  H.~Fukaya {\it et al.} [JLQCD and TWQCD Collaborations],
  Phys.\ Rev.\ D {\bf 83}, 074501 (2011)
  [arXiv:1012.4052 [hep-lat]].
%
  \bibitem{Lattqqn3_high}
S.~Aoki {\it et al.} [PACS-CS Collaboration],
  Phys.\ Rev.\ D {\bf 79}, 034503 (2009)
  [arXiv:0807.1661 [hep-lat]].
  %
\bibitem{Lattqqn3_corr}
C.~T.~H.~Davies {\it et al.} [HPQCD Collaboration],
  Phys.\ Rev.\ D {\bf 100}, no. 3, 034506 (2019)
  [arXiv:1811.04305 [hep-lat]].
  
%
\bibitem{Lam2_latt_ALPHA}
P.~Fritzsch, F.~Knechtli, B.~Leder, M.~Marinkovic, S.~Schaefer, R.~Sommer and F.~Virotta,
  Nucl.\ Phys.\ B {\bf 865}, 397 (2012)
  [arXiv:1205.5380 [hep-lat]].
%
\bibitem{Lam2_latt_other}
B.~Blossier {\it et al.} [ETM Collaboration],
  Phys.\ Rev.\ D {\bf 82}, 034510 (2010)
  [arXiv:1005.5290 [hep-lat]];
  K.~Jansen {\it et al.} [ETM Collaboration],
  JHEP {\bf 1201}, 025 (2012)
  [arXiv:1110.6859 [hep-ph]];
  F.~Karbstein, M.~Wagner and M.~Weber,
  Phys.\ Rev.\ D {\bf 98}, no. 11, 114506 (2018)
  [arXiv:1804.10909 [hep-ph]].
%
\bibitem{LamlattVstatic14} 
F.~Karbstein, A.~Peters and M.~Wagner,
  JHEP {\bf 1409}, 114 (2014)
  doi:10.1007/JHEP09(2014)114
  [arXiv:1407.7503 [hep-ph]].

%
\bibitem{qqlatt_n2}
K.~Cichy, E.~Garcia-Ramos and K.~Jansen,
  JHEP {\bf 1310}, 175 (2013)
  [arXiv:1303.1954 [hep-lat]];
  R.~Baron {\it et al.} [ETM Collaboration],
  JHEP {\bf 1008}, 097 (2010)
  [arXiv:0911.5061 [hep-lat]];
   B.~B.~Brandt, A.~J\"uttner and H.~Wittig,
  JHEP {\bf 1311}, 034 (2013)
  [arXiv:1306.2916 [hep-lat]].
%
  
\bibitem{lattLam0} 
S.~Capitani, M.~L\"uscher, R.~Sommer and H.~Wittig,
  Nucl.\ Phys.\ B {\bf 544}, 669 (1999)
  Erratum: [Nucl.\ Phys.\ B {\bf 582}, 762 (2000)]
  [hep-lat/9810063];
M.~Gockeler, R.~Horsley, A.~C.~Irving, D.~Pleiter, P.~E.~L.~Rakow, G.~Schierholz and H.~Stuben,
  Phys.\ Rev.\ D {\bf 73}, 014513 (2006)
  [hep-ph/0502212];
  N.~Brambilla, X.~Garcia i Tormo, J.~Soto and A.~Vairo,
  Phys.\ Rev.\ Lett.\  {\bf 105}, 212001 (2010)
  Erratum: [Phys.\ Rev.\ Lett.\  {\bf 108}, 269903 (2012)]
  [arXiv:1006.2066 [hep-ph]];
  M.~Kitazawa, T.~Iritani, M.~Asakawa, T.~Hatsuda and H.~Suzuki,
  Phys.\ Rev.\ D {\bf 94}, no. 11, 114512 (2016)
  [arXiv:1610.07810 [hep-lat]];
K.~I.~Ishikawa, I.~Kanamori, Y.~Murakami, A.~Nakamura, M.~Okawa and R.~Ueno,
  JHEP {\bf 1712}, 067 (2017)
  [arXiv:1702.06289 [hep-lat]].
  
%
\bibitem{qqlattn0}
T.~W.~Chiu and T.~H.~Hsieh,
  Nucl.\ Phys.\ B {\bf 673}, 217 (2003)
  [hep-lat/0305016].
%
\bibitem{ss_uulattice}
C.~T.~H.~Davies, C.~McNeile, A.~Bazavov, R.~J.~Dowdall, K.~Hornbostel, G.~P.~Lepage and H.~Trottier,
  PoS ConfinementX, 042 (2012).
%
\bibitem{ss_uuSR}
C.~A.~Dominguez, N.~F.~Nasrallah, R.~Rontsch and K.~Schilcher,
  J. High Energy Phys. {\bf 0805}, 020 (2008).
%
\bibitem{ss_uuSRlast}
S.~Narison and R.~Albuquerque,
  Phys.\ Lett.\ B {\bf 694}, 217 (2010);
R.~M.~Albuquerque, S.~Narison, and M.~Nielsen,
  Phys.\ Lett.\ B {\bf 684}, 236 (2010).
%
\bibitem{qqSRrev18} 
  P.~Gubler and D.~Satow,
  Prog.\ Part.\ Nucl.\ Phys.\  {\bf 106}, 1 (2019)
  [arXiv:1812.00385 [hep-ph]].
%
%
\end{thebibliography}
\end{document}